\documentclass[fleqn,10pt]{wlscirep}
\usepackage[utf8]{inputenc}
\usepackage[T1]{fontenc}

\usepackage{xcolor}
\usepackage{soul}
\usepackage{graphicx}
\usepackage{amsmath,amssymb}
\usepackage{booktabs}
\usepackage{enumitem}
\usepackage{url}
\usepackage{tikz}
\usepackage[section]{placeins}
\usetikzlibrary{arrows.meta,positioning}
\graphicspath{{./}}

\title{Complexity synchronization as a diagnostic and control principle for adaptive systems}

\author[1,*]{Korosh Mahmoodi}
\author[1]{Scott E. Kerick}
\author[1]{Piotr J. Franaszczuk}
\author[1]{David L. Boothe}
\author[2]{Paolo Grigolini}
\author[2]{Bruce J. West}

\affil[1]{U.S. Army Combat Capabilities Development Command Army Research Laboratory, Aberdeen Proving Ground, MD 21005, USA}
\affil[2]{Center for Nonlinear Science, University of North Texas, Denton, TX 76203, USA}
\affil[*]{koroshmahmoodi@gmail.com}

\begin{abstract}
Adaptive systems can exhibit similar levels of performance while relying on fundamentally different internal modes of coordination. Standard metrics such as average cooperation or payoff indicate whether a system succeeds, but do not reveal how coordination is organized across interacting components or which adaptive variables should be targeted when performance fails. Here we propose complexity synchronization (CS)—the synchronization of evolving temporal complexity across coupled variables—as a diagnostic and intervention-guiding principle for adaptive systems.

We test this idea in an adaptive multi-agent system composed of Selfish Algorithm agents interacting in a reduced Predator--Prey model with a Prisoner’s Dilemma--like payoff structure. Temporal complexity is quantified using sliding-window modified diffusion entropy analysis (MDEA), which is sensitive to renewal-like, event-driven restructuring, and detrended fluctuation analysis (DFA), which is sensitive to persistent long-range correlations. CS is defined as the correlation between the resulting time-dependent scaling exponents.

In the high-interaction regime, MDEA-based CS increases with cooperative performance, whereas DFA-based CS captures a distinct persistence-dominated coordination mode. To test whether CS can guide intervention, we selectively weakened learning in the payoff-sharing subsystem and compared three recovery strategies: random/local rescue, equal-budget global rescue, and CS-guided targeted rescue. Random/local and global interventions produced little or only partial recovery, whereas the CS-guided intervention restored cooperation to near-baseline levels and reduced the intervention search space from 15 possible threshold pairs to a single payoff-sharing pair ranked first by the CS-guided search.

These results show that CS can reveal functionally relevant subsystems and provide a principled basis for targeted repair. More broadly, CS offers a general diagnostic and engineering framework for understanding and controlling coordination in biological, social, human--machine, and other adaptive systems.
\end{abstract}

\begin{document}

\flushbottom
\maketitle
\thispagestyle{empty}

\section*{Introduction}

While classical synchronization focuses on the alignment of system states, and criticality emphasizes scaling behavior within individual systems, the present work explores whether coordination in adaptive systems can be understood through synchronization of the evolving complexity of interacting components. The emergence of cooperation in systems composed of self-interested agents is a central problem in the behavioral, biological, and social sciences. The Prisoner’s Dilemma (PD) provides a canonical framework for studying this problem by capturing the tension between individual incentives and collective benefits. In repeated interactions, cooperation can arise despite short-term incentives to defect, making the PD a fundamental model for understanding collective organization in adaptive systems \cite{axelrod1981evolution,nowak2006five}. More generally, the same framework can be extended to competitive settings, where the temporal organization of interaction may be as important as the average outcome itself.

In recent years, complexity science has expanded the concept of synchronization beyond the familiar cases of phase locking and amplitude matching \cite{strogatz2003crucial,pikovsky2001}. This extension is particularly relevant for biological and social systems, where coordination often occurs in signals that are neither periodic nor stationary. Evidence from interpersonal and social neuroscience indicates that coordinated behavior can be accompanied by synchronization across brains, bodies, and autonomic signals during social interaction \cite{dumas2010interbrain,konvalinka2011synchronized}. In such systems, the most informative variable may not be the signal value itself, but the statistical structure of the fluctuations that generate it.

Recent studies \cite{mahmoodi2023complexity,west2023complexity} proposed that interacting organ networks within the human body exhibit a distinct form of synchronization at the level of temporal complexity. In those studies, cardiovascular, respiratory, and electroencephalographic signals displayed synchronized fluctuations of their scaling indices. This phenomenon was termed \emph{complexity synchronization} (CS). In this framework, the relevant quantities are not the raw signals themselves, but the time-dependent scaling exponents that characterize the evolving statistical structure of their dynamics.

A key implication of this perspective is that coordination may be encoded in how complexity changes over time rather than in static similarity or conventional correlation. Systems can exhibit comparable levels of complexity without interacting, just as they can achieve similar performance through qualitatively different internal organizations. By tracking the coevolution of complexity, CS provides a way to reveal how coordination is organized dynamically.

One possible mechanism that can generate temporal complexity is self-organized temporal criticality (SOTC), in which intermittent, scale-free events---referred to as Crucial Events---drive continual restructuring of system dynamics over time \cite{mahmoodi2017self}. In such systems, event-driven fluctuations can produce renewal-like statistics and adaptive flexibility. However, this mechanism is not assumed to be universal. In other adaptive systems, coordination may instead be dominated by persistent memory or long-range correlations. The central question addressed here is therefore not whether one temporal mechanism is always responsible for coordination, but whether CS can diagnose which mode of organization is functionally relevant in a given system.

To investigate this question, we study adaptive multi-agent systems built from Selfish Algorithm (SA) agents \cite{mahmoodi2020selfish}, whose decisions are continuously modified by reinforcement-like feedback from prior payoffs within repeated PD-like interactions. Despite purely self-interested behavior, these agents can generate emergent cooperation and collective organization, providing a minimal framework for studying how coordination arises from adaptive interaction dynamics rather than from externally imposed rules.

A central issue in complex systems is that standard performance measures reveal whether the system succeeds, but not how that success is organized internally. A given level of cooperation or payoff may emerge through qualitatively different temporal regimes, including flexible event-driven restructuring, persistence-dominated dynamics, or mixtures of both. The goal of the present work is therefore not merely to test whether CS covaries with performance, but to determine whether different forms of CS diagnose distinct coordination regimes.

The diagnostic problem is also an intervention problem. In complex networks, failures may arise from changes in a small subset of components, while restoring function may require identifying a low-dimensional intervention target rather than applying global modification. This idea underlies work on network controllability, realistic control of nonlinear network dynamics, and resilience in complex systems \cite{liu2011controllability,cornelius2013realistic,gao2016universal}. In adaptive systems, however, the relevant intervention targets are often not fixed nodes, but evolving internal variables whose organization changes with learning and interaction. This creates the need for a diagnostic field that links coordination structure to candidate intervention targets.

\begin{figure}[tbp]
\centering
\includegraphics[width=0.9\textwidth]{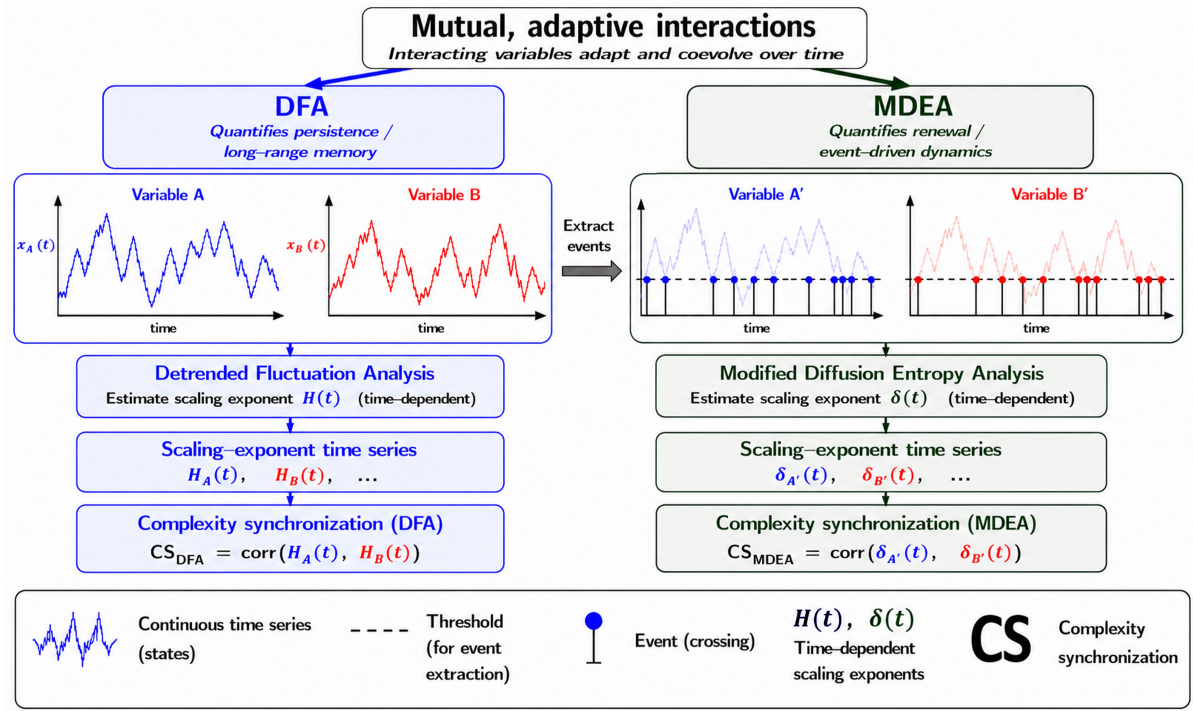}
\caption{
Conceptual framework for complexity synchronization (CS) in interacting systems.
Mutual adaptive interactions generate temporal dynamics that can be characterized through complementary scaling measures.
For two interacting variables $i$ and $j$ (e.g., adaptive thresholds from two agents), detrended fluctuation analysis (DFA) quantifies persistence and long-range memory, producing scaling time series $H_i(t)$ and $H_j(t)$ whose synchronization defines $CS_{\mathrm{DFA}}$.
Modified diffusion entropy analysis (MDEA) quantifies renewal-like, event-driven dynamics. For visual clarity, the MDEA panel illustrates event extraction using a single threshold, with events marked at threshold crossings. In the actual analyses reported throughout the paper, events are not defined from a single threshold. Instead, multiple equally spaced thresholds (stripes) are used, and events are defined as the times at which the signal crosses from one stripe to another. The resulting event sequence is then used to estimate the scaling time series $\delta_i(t)$ and $\delta_j(t)$, whose synchronization defines $CS_{\mathrm{MDEA}}$.
}
\label{concept_pipeline}
\end{figure}

Our central hypothesis is that complexity synchronization provides such a diagnostic field. As summarized in Fig.~\ref{concept_pipeline}, interactions among adaptive components generate temporal dynamics whose complexity can be quantified by tracking local scaling exponents over time. For simplicity, the MDEA schematic in Fig.~\ref{concept_pipeline} shows event extraction using a single threshold. In the actual analyses, however, events are obtained using multiple equally spaced stripes, and each event corresponds to a crossing from one stripe to another. CS is then defined as the correlation between these time-dependent scaling signals across interacting variables. In this framework, CS does not measure performance directly; rather, it characterizes how the internal organization of the system evolves and identifies adaptive subsystems that may be functionally important.

To characterize this organization, we use two complementary measures of temporal complexity. Modified diffusion entropy analysis (MDEA) is sensitive to renewal-like, event-driven restructuring, whereas detrended fluctuation analysis (DFA) emphasizes persistent long-range correlations. Together, these methods provide distinct but complementary views of adaptive dynamics and, as illustrated in Fig.~\ref{concept_pipeline}, define corresponding measures of CS. Comparing MDEA- and DFA-based CS therefore allows us to determine which temporal mode of organization is most closely associated with system performance and most informative for guiding intervention.

In the present paper, we test this framework in a reduced Predator--Prey model and examine how CS relates to cooperation, perturbation, and recovery. In this model, increasing cooperation is associated with stronger MDEA-based CS, consistent with a shift toward more renewal-like, event-driven dynamics. However, this result should be interpreted as model-specific rather than universal. In other adaptive systems, cooperation or performance could instead be associated with synchronized memory-dominated dynamics, in which case DFA-based CS might exhibit a positive relationship with performance. The broader conclusion is therefore not that one scaling framework is always superior, but that comparing complementary forms of CS can reveal the temporal organization of coordination and guide targeted intervention in adaptive systems.

\section*{Basic concepts and models}\label{Model}

This section introduces the minimal set of concepts required to interpret the results. Rather than providing a full review, we focus on how temporal complexity, adaptive interaction, and synchronization are defined and measured in the present work.

\subsection*{Temporal complexity and event-driven dynamics}

In many adaptive systems, behavior is not governed by a characteristic time scale but by intermittent fluctuations spanning multiple scales. Such dynamics are often associated with event-driven processes in which the system undergoes irregular reorganizations. These events, referred to as \emph{Crucial Events}, can be characterized statistically by heavy-tailed waiting-time probability density distributions,
\[
\psi(\tau)\propto \tau^{-\mu}, \quad 1<\mu<3,
\]
where $\tau$ is the time interval between two consecutive events. 
Rather than describing the system through averages or instantaneous values, this framework emphasizes the timing and structure of these events. When such dynamics are present, the system exhibits scale-free temporal organization, often associated with adaptive flexibility and responsiveness.

\subsection*{Measuring complexity}

To quantify complexity, we analyze time series using two complementary approaches.

Modified diffusion entropy analysis (MDEA) extracts scaling $\delta$ from event sequences derived from the original signals. By converting the signal into a sequence of events via stripe crossing, MDEA captures the statistical structure of inter-event intervals and is therefore sensitive to renewal-like dynamics. Although the conceptual illustration in Fig.~\ref{concept_pipeline} uses a single threshold to visualize the event-generation process, all reported MDEA analyses use multiple equally spaced stripes. Events are defined as the times at which the signal crosses from one stripe to another, generating the event sequence used for scaling estimation.

In contrast, detrended fluctuation analysis (DFA) estimates scaling $H$ based on correlations within the signal itself. DFA emphasizes persistent long-range correlations and is therefore more sensitive to memory-dominated dynamics.

These two measures provide different perspectives on complexity. Details of MDEA and DFA are provided in Supplementary Sections S4 and S5. In the present study, their contrast is not a limitation but a diagnostic tool for distinguishing coordination regimes.

\subsection*{Complexity synchronization}

Complexity synchronization (CS) refers to the alignment of the complexity measure across interacting components. Operationally, for different signals, we compute scaling exponents on overlapping windows and define CS as the correlation between these time-dependent scaling signals ($\delta(t)$  or $H(t)$ ).

This definition differs from conventional synchronization, which focuses on alignment of amplitudes or phases of signals. Here, synchronization occurs at the level of evolving statistical structure. As a result, two signals can exhibit strong CS even when their raw time series are weakly correlated.

Importantly, CS captures the \emph{temporal co-variation} of complexity, rather than its static value. This distinction allows CS to detect coordinated dynamics that unfold over time, providing a direct measure of how interaction is organized. Unlike measures based solely on similarity, CS does not saturate once systems reach comparable complexity; instead, it quantifies the degree to which their internal organizations evolve together, enabling a continuous assessment of coordination.

\subsection*{Relation to complexity matching}

It is useful to distinguish CS from complexity matching. Complexity matching refers to optimal information transfer when interacting systems have similar levels of complexity, typically quantified through comparable scaling exponents. This idea is closely related to \emph{Ashby's Law of Requisite Variety}, which states that effective control requires a controller with at least the same level of complexity as the system being controlled. In this sense, matching complexity defines the conditions under which systems can efficiently exchange information or exert control.

However, similarity in complexity alone does not establish interaction. Two systems may exhibit comparable scaling properties due to shared constraints, common environments, or intrinsic dynamics, without being functionally coupled. Therefore, complexity matching provides a \emph{necessary but not sufficient} condition for interaction: it defines the \emph{capability} for efficient information exchange, but does not indicate whether such exchange is actually taking place.

By contrast, CS concerns the \emph{co-evolution} of complexity over time. By tracking whether fluctuations in complexity are temporally coordinated across systems, CS provides evidence that the potential for interaction is actively realized. In this way, complexity matching defines a compatibility condition, whereas CS reveals the dynamical organization of coordination itself.

In the present work, we focus on synchronization of changing complexity rather than on static similarity, because adaptive systems continuously reorganize their internal dynamics. This perspective allows us to move beyond structural compatibility and directly quantify how coordination emerges and evolves in time.

\subsection*{Selfish Algorithm: probabilistic decision with adaptive thresholds}

We model adaptive decision-making using Selfish Algorithm (SA) agents \cite{mahmoodi2020selfish,mahmoodi2019emergence}. The core mechanism is a single adaptive threshold that governs a binary decision.

At each time step, an agent compares its internal threshold $X(t) \in [0,1]$ to a uniformly distributed random number $u \sim \mathcal{U}(0,1)$. The realized action is given by
\[
s(t) =
\begin{cases}
1, & \text{if } u > X(t), \\
0, & \text{otherwise}.
\end{cases}
\]
In this way, the threshold defines the probability of selecting one of two alternative actions (e.g., trust vs.\ not trust), while preserving stochastic variability.

After each decision, the threshold is updated through payoff-driven reinforcement. Actions that lead to higher payoff shift the threshold to increase the likelihood of repeating that decision, whereas unfavorable outcomes shift it in the opposite direction. As a result, the threshold evolves as a compact internal memory encoding past experience.

This simple mechanism generates adaptive, history-dependent behavior while maintaining variability at each decision step.

\subsection*{From single-threshold decisions to adaptive multi-agent interaction}

Realistic adaptive behavior requires multiple coupled decisions. In the SA framework, each agent therefore carries a set of thresholds, each governing a distinct behavioral dimension. In the present study, these correspond to information sharing $I(t)$, trust $T(t)$, and payoff sharing $P(t)$.

Each threshold independently generates a probabilistic binary decision through comparison with a random number, while all thresholds are jointly updated through shared payoff feedback. This creates a set of interacting internal variables whose dynamics encode the history of adaptation.

\begin{figure}[tbp]
\centering
\includegraphics[width=0.5\textwidth]{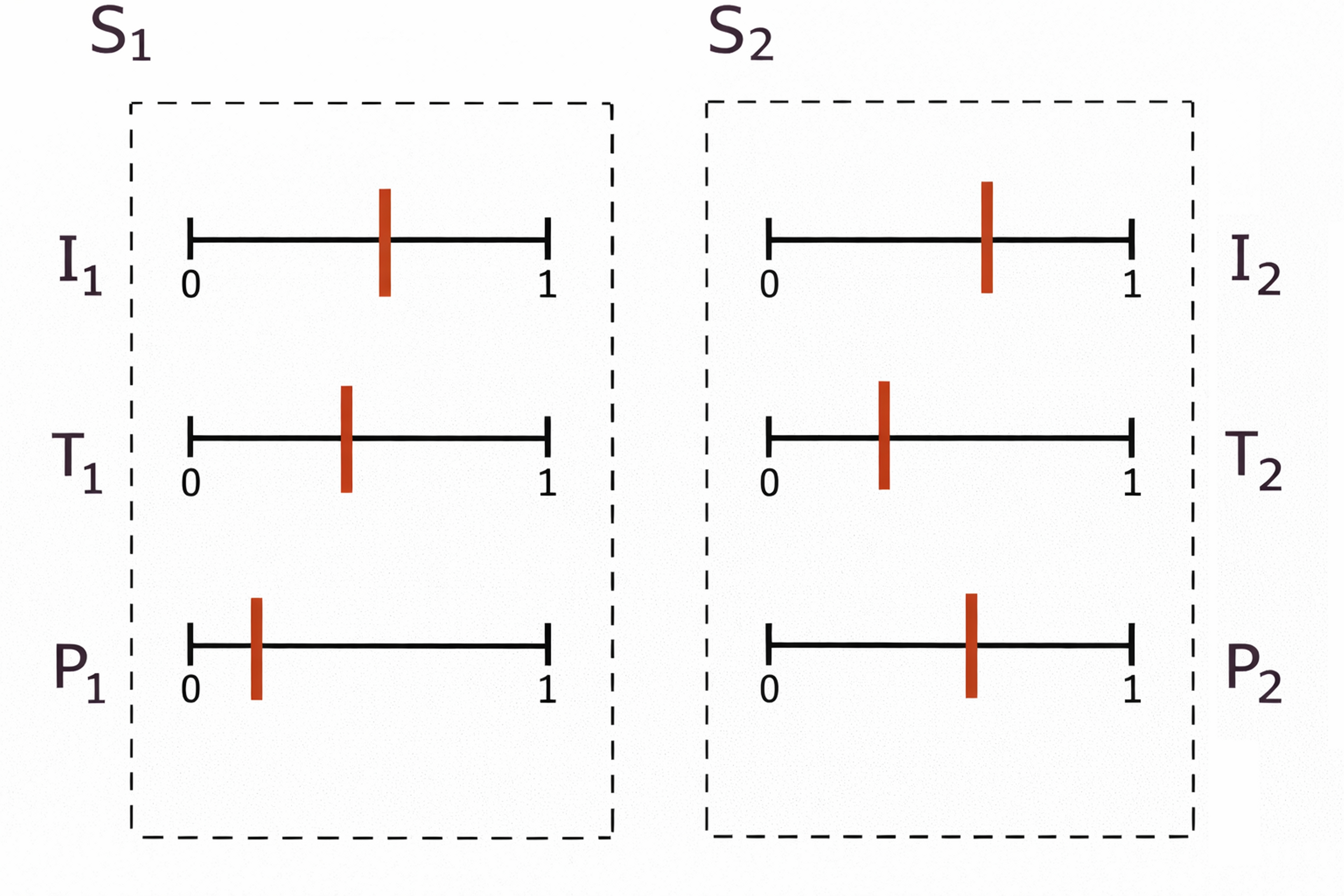}
\caption{Adaptive decision structure of the two interacting Selfish Algorithm  (the two predator agents of the model). Each agent ($S_1$ and $S_2$) is defined by three thresholds governing information sharing ($I$), trust ($T$), and payoff sharing ($P$), with values in $[0,1]$ that determine action probabilities through stochastic sampling. The red bars indicate the current threshold values, which are continuously updated via payoff-dependent reinforcement. This multi-threshold representation forms the basis for the coupled dynamics analyzed through CS.}
\label{fig:sa_decision}
\end{figure}

The adaptive decision architecture of the two Selfish Algorithm agents is illustrated in Fig.~\ref{fig:sa_decision}.

\begin{figure}[tbp]
\centering
\includegraphics[width=0.5\textwidth]{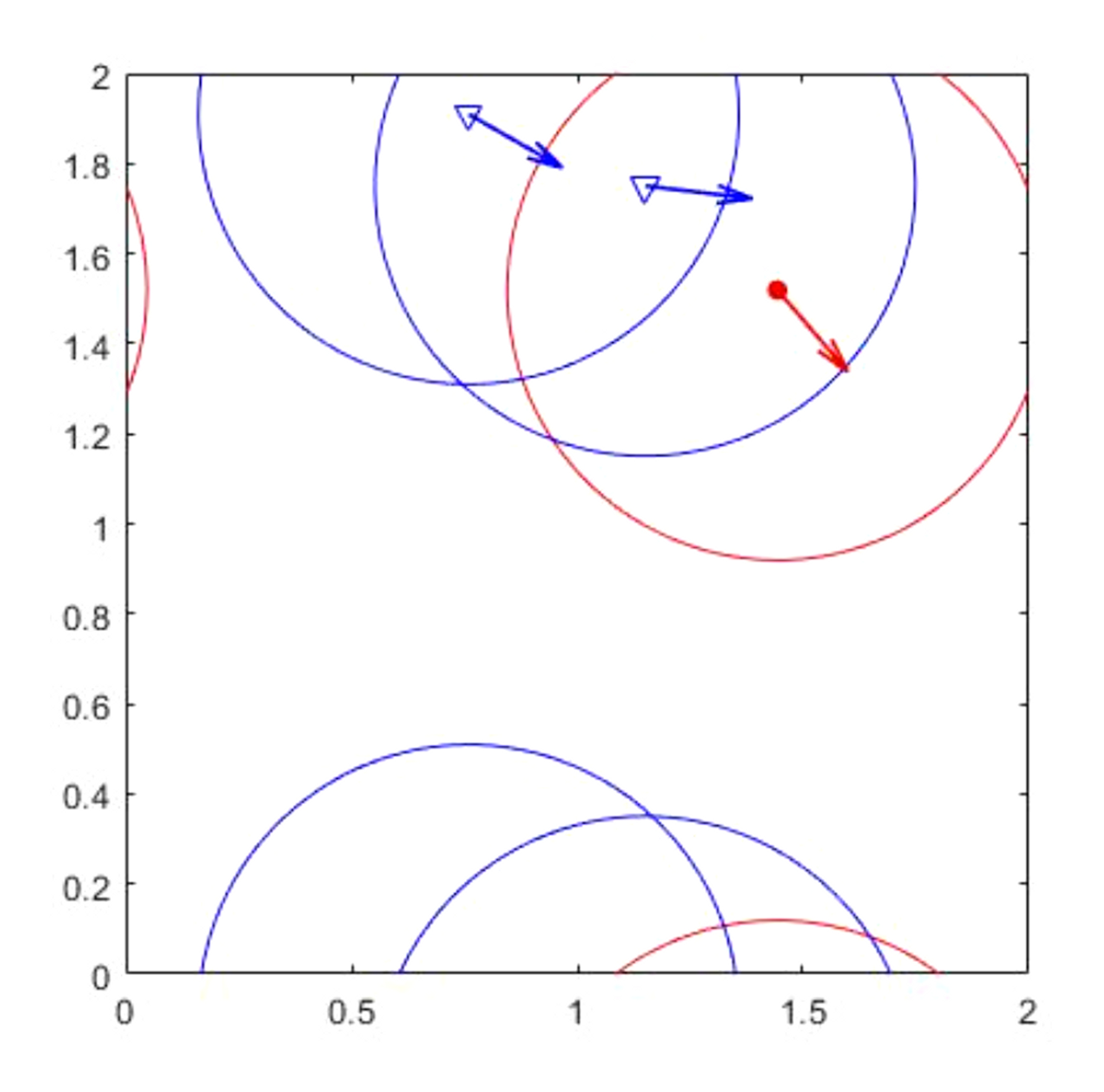}
\caption{Representative snapshot of the spatial interaction dynamics in the Predator--Prey environment. Agents move on a two-dimensional domain with periodic boundary conditions, where the Prey (red) and Predators (blue) interact within finite sensing radii $r_S$ (circles). Arrows indicate instantaneous heading directions, and interactions are determined by local geometric proximity and predicted motion. This spatial framework defines the information available to each agent and underlies the reduced model used for the analysis of adaptive decision dynamics and CS.}
\label{fig:sa_environment}
\end{figure}

To study how such adaptive decision processes generate coordination, we use a reduced Predator--Prey model (Fig.~\ref{fig:sa_environment}). The system consists of two Predator agents and one Prey agent evolving on a periodic two-dimensional domain. Each Predator carries three adaptive thresholds $I(t)$, $T(t)$, and $P(t)$, which determine how it shares information, trusts its partner, and allocates payoff.

The Prey acts as a dynamic and partially unpredictable environment, updating its motion in response to the Predators. This introduces nonstationarity and ensures that agents must continuously adapt rather than converge to fixed strategies.

Coupling between Predators arises primarily through trust-mediated interaction, where one agent may adopt the partner’s estimate of the Prey. As a result, the adaptive thresholds of the two agents become dynamically linked through both shared environmental feedback and direct interaction.

This minimal structure preserves the essential ingredients needed to study coordination: stochastic decision-making, adaptive learning, interaction-driven coupling, and a changing environment.

\subsection*{Prisoner’s Dilemma structure and meaningful cooperation}

To ensure that cooperation is nontrivial, the interaction between agents is embedded in a Prisoner’s Dilemma (PD)-like payoff structure.

In this setting, mutual cooperation (both agents sharing payoff) yields a beneficial outcome for both agents, while unilateral defection provides a higher immediate reward to the defector at the expense of the other agent. Mutual defection results in a poorer outcome for both.

This structure creates a fundamental tension between short-term individual gain and long-term collective benefit. Importantly, cooperation is not imposed by the model but must emerge through adaptive learning despite the persistent incentive to defect.

The PD-like payoff therefore ensures that observed cooperative behavior reflects genuine coordination rather than trivial alignment or externally enforced rules. It provides a minimal and principled framework in which adaptive agents must learn to balance exploitation and cooperation through repeated interaction.

Within this context, cooperation corresponds to sustained mutual payoff-sharing events, while the underlying dynamics are encoded in the evolution of the adaptive thresholds. This allows us to investigate not only whether cooperation emerges, but how it is organized in time.

\FloatBarrier

\section*{Methods}

\subsection*{Reduced Predator--Prey model}

We study a minimal adaptive multi-agent system consisting of two Predator agents and one Prey agent moving on a two-dimensional domain with periodic boundary conditions (Fig.~\ref{fig:sa_environment}). The Predators represent the potentially cooperative elements of the system, while the Prey provides a dynamic and partially unpredictable environment that continuously reshapes the interaction.

\paragraph*{Environment and sensing.}
Each agent interacts with its surroundings through a finite sensing (vision) radius $r_S$, defining the set of neighboring agents it can detect at each time step. For parameter sweeps, we use the dimensionless sensing-radius ratio $R=r_S/L$, where $L$ is the side length of the square domain; thus $r_S=R\times L$. Based on this local information, agents estimate target directions and update their motion on the periodic domain.

\paragraph*{Adaptive decision dynamics.}
Each Predator is characterized by three internal adaptive variables associated with information sharing, trust, and payoff sharing. These variables encode behavioral propensities and evolve over time through payoff-driven learning. At every time step, they generate probabilistic binary decisions, allowing agents to flexibly switch between alternative actions.

\paragraph*{Interaction and information exchange.}
Predators estimate the Prey’s direction by projecting its motion based on local observations. Due to the Prey’s deceptive behavior, this estimate is uncertain and can take one of two possible forms. When a Predator relies on its partner, it replaces its estimate with the complementary estimate inferred by the other Predator. This complementary exchange allows the pair to span the uncertainty in the Prey’s trajectory and constitutes the primary coupling mechanism.

The Prey, in turn, anticipates the Predators’ motion and selects between two symmetric evasive directions at each step. This stochastic two-option mechanism introduces intrinsic uncertainty and maintains a nonstationary environment.

\paragraph*{Payoff and cooperation.}
Successful capture events—determined by proximity within an interaction radius—can trigger
a payoff-sharing interaction between the Predators. In this stage, agents face a Prisoner’s
Dilemma--like trade-off between cooperation and defection. In the implementation used here,
the payoff-sharing stage is entered only if at least one Predator successfully captures the
Prey and at least one trust decision is realized. Mutual cooperation is then recorded when
both Predators choose to share payoff, i.e.,
\[
s^{(P)}_1(t)=s^{(P)}_2(t)=1.
\]
The cooperation rate $C_r$ is defined as the running fraction of these events (i.e., both agents decide to share payoff) over time.

This reduced model combines local sensing, adaptive internal dynamics, complementary information exchange, and a dynamically deceptive environment in a minimal setting that allows the emergence and organization of cooperation to be studied. Full implementation details are provided in the Supplementary Information.

\subsection*{Complexity analysis}

The empirical signals analyzed in the present study are the threshold time series
\[
I_1,\ T_1,\ P_1,\ I_2,\ T_2,\ P_2,
\]
corresponding to information-sharing, trust, and payoff-sharing thresholds for the two Predators. We analyze the temporal complexity of these signals using modified diffusion entropy analysis (MDEA). MDEA converts each threshold trajectory into an event sequence by stripe crossing. Events are defined as crossings of equally spaced amplitude thresholds (“stripes”) applied to the signal. Smaller stripe sizes increase the density of detected events and, in the limit, approach noise-dominated regimes, whereas larger stripe sizes capture coarser, more structured fluctuations.

For all reported MDEA analyses, local scaling exponents were estimated using sliding windows of length $10^4$ samples with 75\% overlap after removal of the first 25\% of each simulation as burn-in. Three representative stripe sizes ($0.1$, $0.01$, and $0.001$) were examined. Within each window, MDEA was applied to the stripe-crossing event sequence to obtain a local scaling exponent $\delta(t)$. Repeating this procedure across overlapping windows generated the time-dependent scaling signals used for complexity-synchronization analysis.

For comparison, we also estimate scaling exponents $H(t)$ using detrended fluctuation analysis (DFA), which analyzes correlations directly in the original time series rather than through event sequences, and is therefore more sensitive to persistent long-range memory.

CS is then defined as the correlation between the resulting scaling-exponent time series. In this way, CS measures synchronization of changing statistical structure rather than synchronization of raw amplitudes.

\subsection*{Renewal-support analysis}

To provide supporting mechanistic evidence for the event-driven interpretation of the MDEA results, we also analyzed the renewal properties of stripe-crossing events. These analyses were performed as representative tests on the information-sharing threshold of Predator 1, $I_1$, rather than on all threshold variables or on the CS pairs directly. The goal was not to define CS, but to examine whether the event sequence extracted from an adaptive threshold exhibits renewal-like organization in the low- and high-cooperation regimes.

For the survival-function aging analysis shown in Fig.~\ref{fig:supp_renewal_test}, the first half of each simulation was discarded, and a single window of length $10^4$ samples was selected from the remaining $I_1$ trajectory. Stripe-crossing events were extracted using the same stripe sizes used in the MDEA analysis ($0.1$, $0.01$, and $0.001$). The inter-event interval $\tau$ was defined as the number of samples between successive crossings from one stripe to another. The aging time was fixed at $t_a=100$, and the aged survival function was compared with a shuffled-aged control obtained by randomly permuting the inter-event intervals before applying the same aging procedure.

For the inter-event-interval autocorrelation analysis shown in Fig.~\ref{fig:supp_autotau}, the first half of the $I_1$ trajectory was again discarded. Autocorrelations were then computed from the $\tau$ sequences extracted from 10 windows of length $10^4$ samples with 75\% overlap. The plotted curves show the mean autocorrelation across these windows, with variability shown as the window-to-window standard deviation. Because the autocorrelation functions can oscillate around zero, a power-law fit to the upper envelope of the absolute autocorrelation was used as a compact summary of the long-lag decay rate; more negative envelope slopes indicate faster decay of inter-event memory.

\subsection*{Simulation summary}

The simulations reported here sweep the dimensionless sensing-radius ratio $R=r_S/L$ over 21 values, use 10 ensembles per condition, and evaluate CS for three representative stripe sizes. Each run contains $10^6$ trials, with a burn-in period removed before analysis. The learning increment is fixed at $\Delta_S=0.1$, and a small stochastic term is included in the threshold update rule. Full numerical details, ordinary-correlation control analyses, and renewal-related supporting analyses are provided in the Supplementary Information.

\subsection*{Perturbation and rescue protocol}

To test whether the CS field can guide intervention, we applied perturbations to the learning increments of the adaptive thresholds. The threshold order used throughout the analysis is
\[
(I_1,T_1,P_1,I_2,T_2,P_2).
\]

The baseline learning vector is
\[
\boldsymbol{\Delta}_{\mathrm{base}}=0.1\,(1,1,1,1,1,1),
\]
whereas the damaged condition reduces the payoff-sharing learning increments for both agents,
\[
\boldsymbol{\Delta}_{\mathrm{damage}}=(0.1,0.1,0.025,0.1,0.1,0.025).
\]
Thus, the perturbation selectively weakens the adaptive channel associated with the payoff-sharing thresholds $P_1$ and $P_2$, while leaving the information-sharing and trust-related learning increments unchanged.

We then compared three rescue strategies. In the random/local strategy, the same local intervention budget is applied to a non-targeted threshold pair. In the equal-budget global strategy, the rescue budget is distributed across all six thresholds. In the CS-guided targeted strategy, the rescue is applied to the threshold pair identified from the CS network as the relevant payoff-sharing subsystem, $P_1$--$P_2$.

Random/local rescue tests whether an arbitrary local intervention can restore function, whereas equal-budget global rescue tests whether distributing the same total intervention budget across all adaptive variables is sufficient. These controls distinguish targeted, diagnosis-driven repair from non-specific interventions.

The rescue experiments assume that the nominal learning increments of the healthy system are known and that restoring damaged parameters toward their baseline values is a reasonable intervention strategy. The central challenge is therefore not to determine the direction of adjustment, but to identify which subset of adaptive variables should be targeted first.

Rescue strength $s\in[0,1]$ interpolates continuously between the damaged and baseline learning increments. For a damaged learning increment $\Delta_{\mathrm{damage}}$ and its corresponding baseline value $\Delta_{\mathrm{base}}$, the rescued value is defined as
\[
\Delta(s)
=
\Delta_{\mathrm{damage}}
+
s\left(
\Delta_{\mathrm{base}}
-
\Delta_{\mathrm{damage}}
\right),
\qquad 0 \le s \le 1.
\]
Accordingly, $s=0$ corresponds to the damaged system, and $s=1$ corresponds to full restoration of the selected learning increment(s).

For targeted rescue, this interpolation is applied only to the damaged payoff-sharing pair $P_1$--$P_2$. For random/local rescue, the same local intervention magnitude is applied to a non-targeted threshold pair, while the damaged payoff-sharing pair remains weakened. For equal-budget global rescue, the same total rescue budget is distributed evenly across all six adaptive thresholds, so that the intervention is diffuse rather than localized.

Because the system contains six adaptive thresholds, 15 distinct threshold pairs can be tested as candidate intervention targets. In Fig.~\ref{fig:search_efficiency}, the horizontal axis indicates the cumulative number of tested pairs. The CS-guided search ranks candidate pairs according to the strength of their baseline CS links, under the assumption that strongly synchronized pairs are more likely to represent functionally relevant subsystems, whereas the random search samples candidate pairs without using CS information. In each test, the selected threshold pair is restored to its baseline learning increment while all remaining thresholds retain their damaged values.

All rescue analyses use $10^6$ trials and ENS = 10 independent stochastic ensembles, with distinct random seeds for each ensemble and condition.

\begin{figure}[tbp]
\centering
\includegraphics[width=0.72\textwidth]{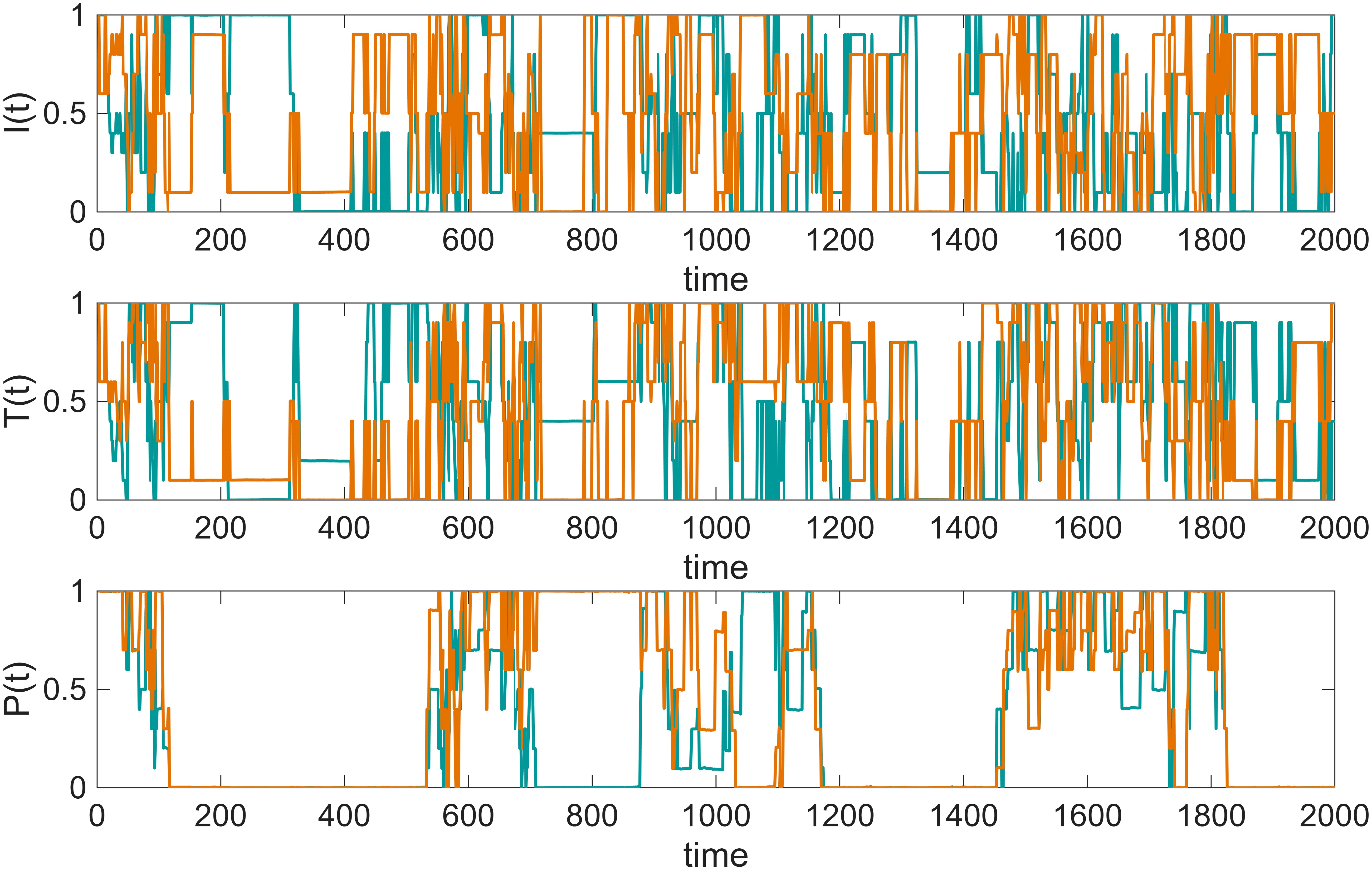}
\caption{Representative short time series of the three adaptive threshold variables, information sharing $I(t)$, trust $T(t)$, and payoff sharing $P(t)$, over 2000 trials (out of $10^6$). The two colored traces correspond to the two Predator agents. These threshold dynamics form the raw signals from which local scaling exponents are extracted using MDEA and DFA.}
\label{fig:mdea_example}
\end{figure}

A representative segment of the adaptive threshold dynamics used as input to the scaling analyses is shown in Fig.~\ref{fig:mdea_example}.

\FloatBarrier

\section*{Results}\label{sec2}

\subsection*{Learning improves cooperation and payoff}

\begin{figure*}[tbp]
\centering
\includegraphics[width=1\textwidth]{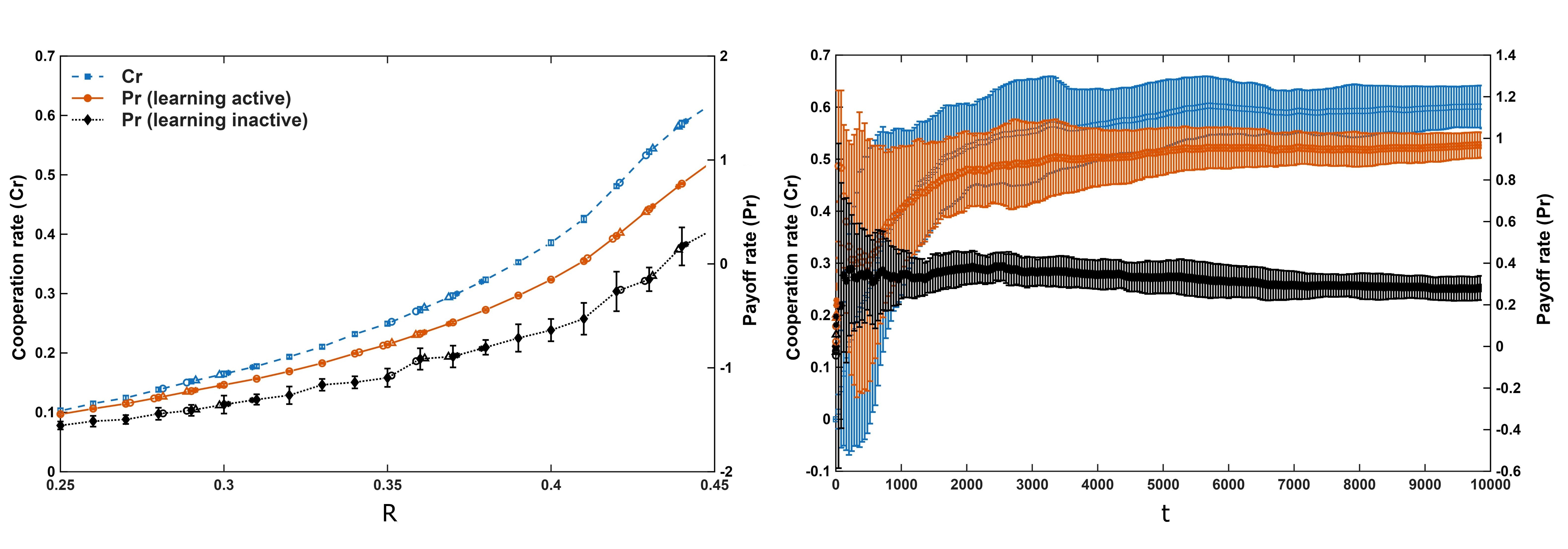}
\caption{Performance validation of the Predator--Prey model. Left panel: Cooperation rate $C_r$ and payoff rate $P_r$ (at $t=10^6$) are shown as functions of the sensing radius ratio $R$ for simulations with learning and without learning. Right panel: time evolution of the same quantities for $R = 0.45$.}
\label{fig:performance_validation}
\end{figure*}

Before turning to CS, we first verify that adaptive learning changes the behavioral outcome of the model. Figure~\ref{fig:performance_validation} shows that learning increases both the cooperation rate $C_r$, which is the cumulative cooperation rate, and the payoff rate $P_r$, which is the cumulative payoff rate of the predators, relative to the non-learning baseline across the dimensionless sensing-radius ratio $R=r_S/L$ (equivalently, $r_S=R\times L$; left panel), and also improves the time-resolved trajectories of these quantities in representative runs (right panel). In both panels, the curves show the average over ten simulations. This establishes that the adaptive threshold dynamics generate a meaningful cooperative regime whose temporal organization can then be analyzed using complexity-based measures.

\FloatBarrier

\subsection*{Scaling estimation from adaptive threshold dynamics}

\begin{figure}[tbp]
\centering
\includegraphics[width=\textwidth]{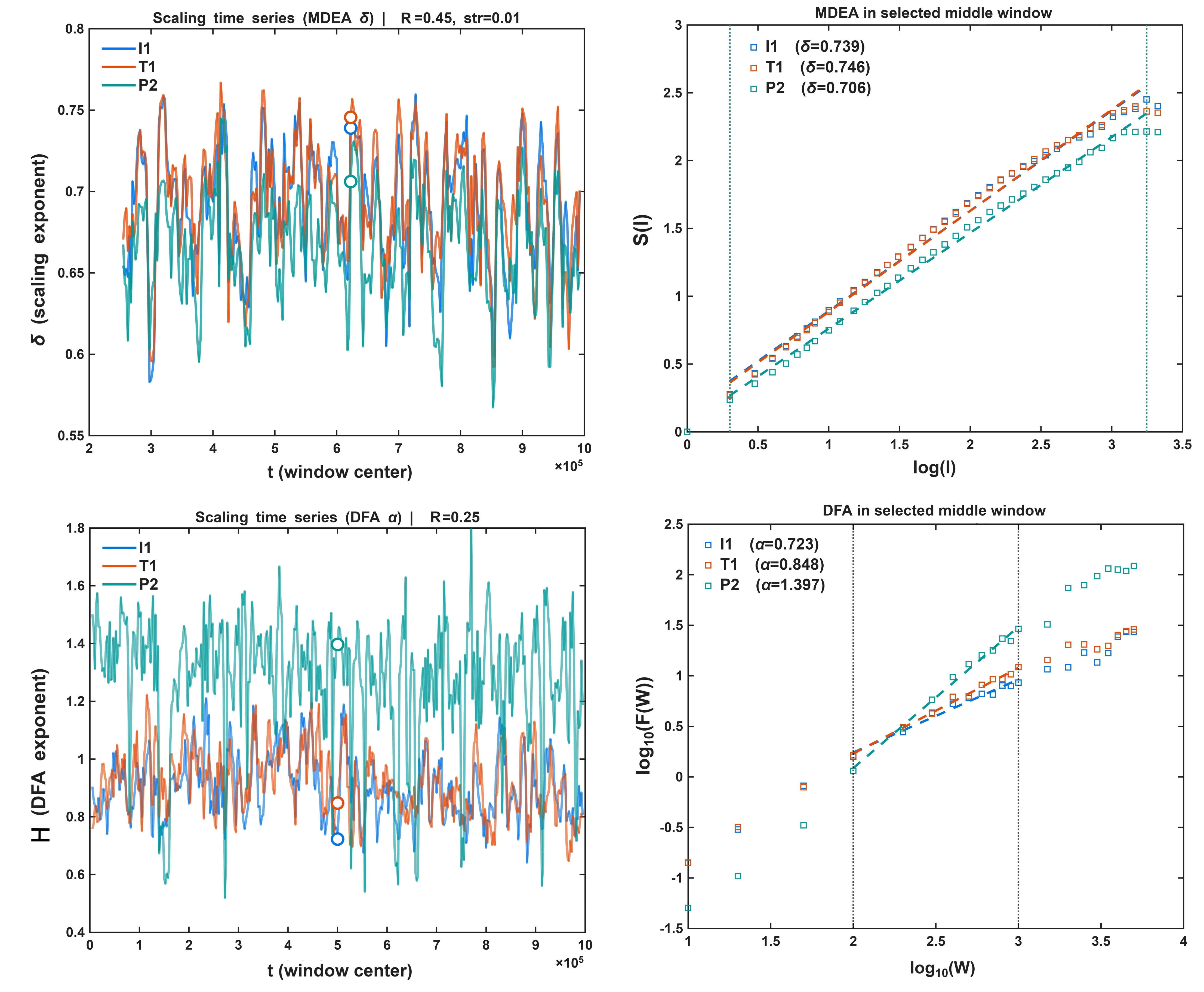}
\caption{Example of local scaling estimation from the adaptive threshold dynamics. Left panels: scaling-exponent time series obtained from sliding windows using MDEA (top) and DFA (bottom). Right panels: the corresponding scaling plots for a selected window, marked by circles, in the left panels. In the top-right panel, the entropy $S(l)$ is plotted against $\log(l)$, where $l$ is the length of the moving window in the MDEA; in the bottom-right panel, the fluctuation function is plotted on log--log axes, where $W$ is the length of the moving window in the DFA. The fitted slopes define the local scaling exponents used throughout the paper. For the details of MDEA and DFA, see Supplementary Sections S4 and S5, respectively.}
\label{fig:scaling_estimation}
\end{figure}

The first step in the complexity analysis is to estimate local scaling exponents from the threshold signals. Figure~\ref{fig:scaling_estimation} illustrates this procedure for both MDEA and DFA. The left panels show the resulting time-dependent scaling exponents obtained from sliding windows, whereas the right panels show representative scaling plots from one selected window. These local scaling signals form the basis for the CS measures used throughout the remainder of the Results section.

\FloatBarrier

\subsection*{Cooperation and CS}

The main quantitative result of the study is the relationship between cooperative performance and CS. When CS is evaluated using MDEA, the cooperation rate $C_r$ increases with synchronization of scaling dynamics (Fig.~7). In contrast, the same comparison based on DFA yields the opposite trend (Fig.~\ref{fig:cc_cs_dfa}).

\begin{figure*}[tbp]
\centering
\includegraphics[width=1.1\textwidth]{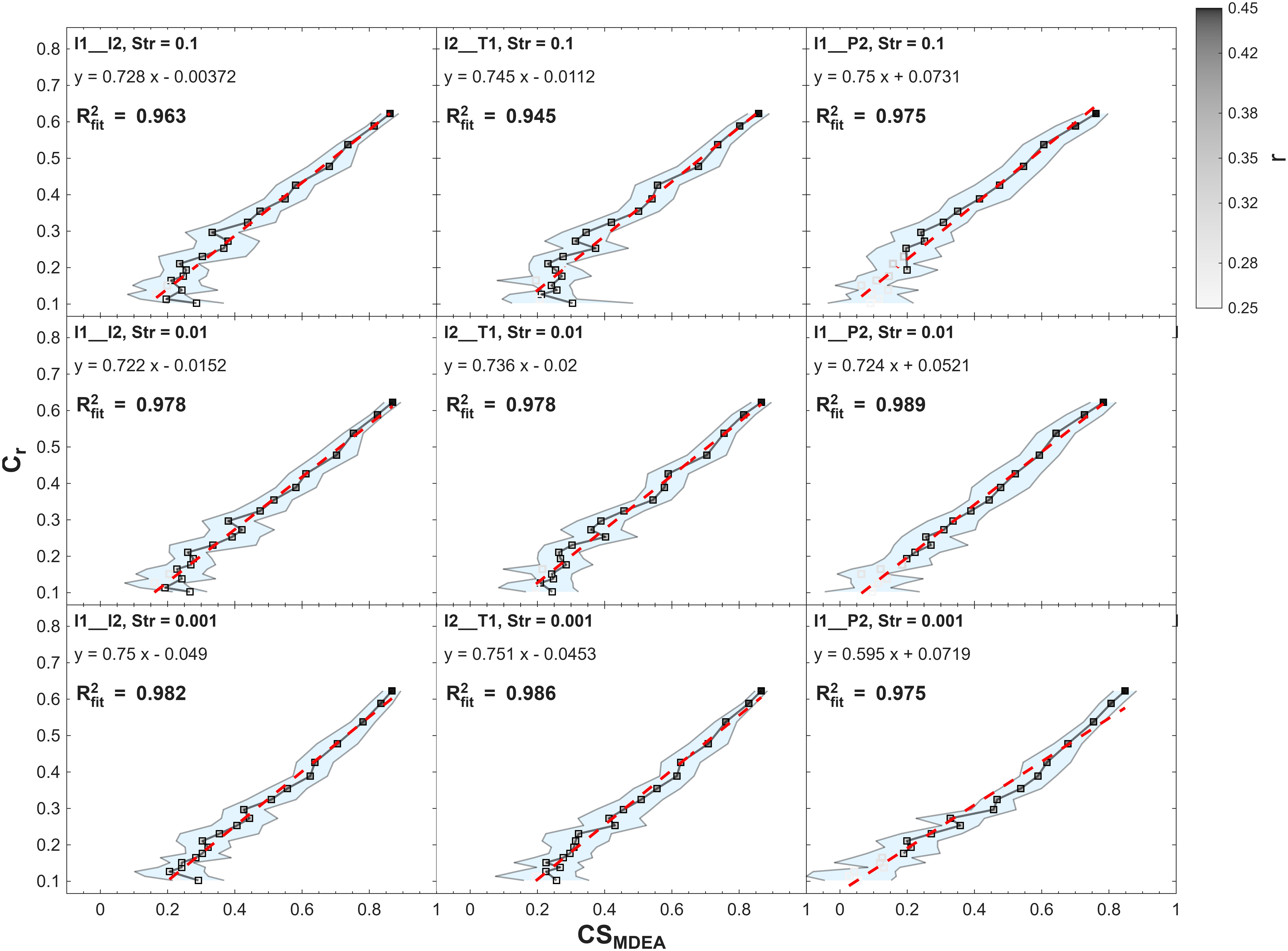}
\caption{Relationship between  cooperation rate $C_r$ and MDEA-based CS. Columns correspond to representative threshold pairs ($I_1$--$I_2$, $I_2$--$T_1$, and $I_1$--$P_2$), and rows correspond to stripe sizes used in MDEA ($\mathrm{Str} = 0.1, 0.01, 0.001$). Each marker represents a simulation at a different sensing ratio $R \in [0.25, 0.45]$, with grayscale indicating the $R$ value (color bar). Filled squares denote statistically significant CS values ($p < 0.05$), whereas open squares indicate non-significant values. Red dashed lines show linear fits, with corresponding regression equations and $R^2_{fit}$ values reported in each panel. Shaded bands indicate variability across realizations (mean $\pm$ standard deviation).}
\label{fig:cc_cs_mdea}
\end{figure*}

\begin{figure*}[tbp]
\centering
\includegraphics[width=1.1\textwidth]{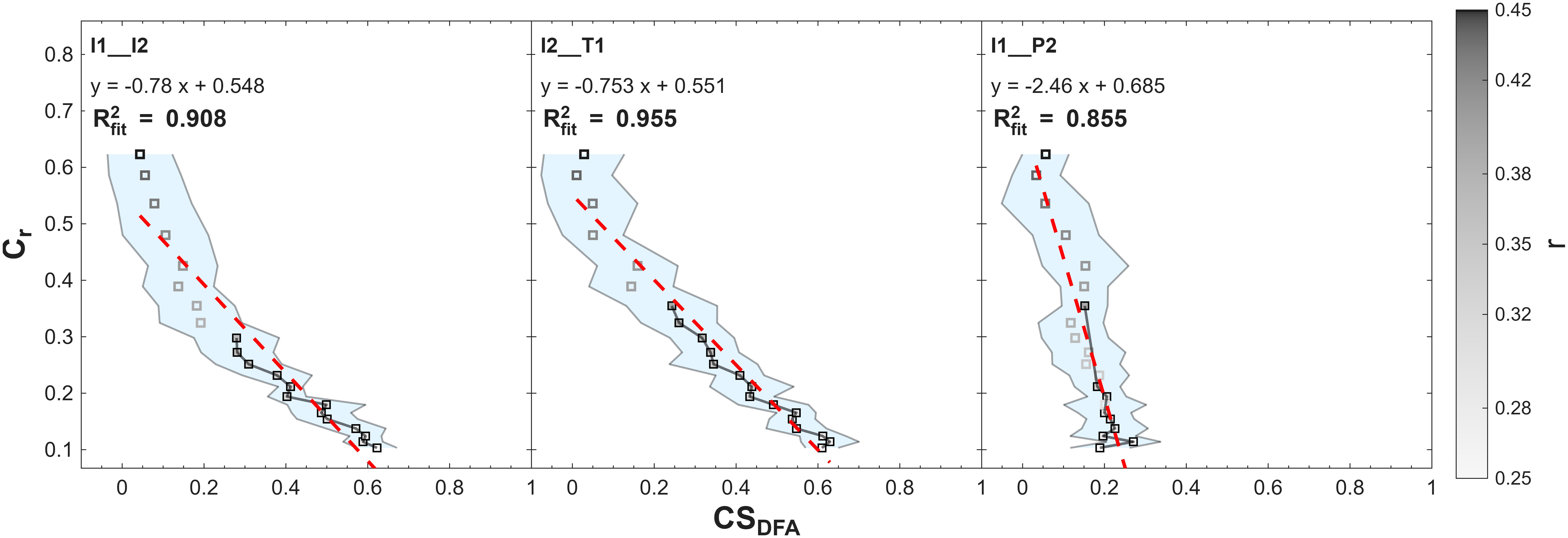}
\caption{Relationship between  cooperation rate $C_r$ and DFA-based CS. Each panel corresponds to a representative threshold pair ($I_1$--$I_2$, $I_2$--$T_1$, and $I_1$--$P_2$), with scaling exponents estimated using detrended fluctuation analysis (DFA). Each marker represents a simulation at a different sensing ratio $R \in [0.25, 0.45]$, with grayscale indicating the $R$ value (color bar). Filled squares denote statistically significant CS values ($p < 0.05$), whereas open squares indicate non-significant values. Red dashed lines show linear fits, with corresponding regression equations and $R^2_{fit}$ values reported in each panel. Shaded bands indicate variability across realizations (mean $\pm$ standard deviation).}
\label{fig:cc_cs_dfa}
\end{figure*}

Taken together, the results for the present reduced Predator--Prey model support the empirical relations
\[
C_r \propto CS_{\mathrm{MDEA}}
\]
but
\[
C_r \propto -\,CS_{\mathrm{DFA}}
\]

The consistency of these trends across multiple threshold pairs, stripe sizes, and interaction regimes strengthens the inference that, within this model, the observed contrast is structural rather than anecdotal. We emphasize, however, that the sign of the DFA-based relation need not be universal across all adaptive systems. In other models, cooperation could in principle be supported by synchronized persistence, stable memory, or common slow environmental forcing, in which case a positive relation between $C_r$ and $CS_{\mathrm{DFA}}$ could emerge. The distinctive result here is therefore not that DFA-based CS must always oppose cooperation, but that in the present Predator--Prey model cooperative organization is more strongly aligned with event-driven synchronization than with synchronization of persistent long-range memory. The ordinary Pearson-correlation control analysis is provided in Supplementary Fig.~\ref{fig:supp_rawcorr}.

We therefore interpret the contrast between MDEA-based and DFA-based CS not simply as a difference in predictive direction, but as a diagnostic signature of distinct coordination modes. In this sense, the point of the present analysis is not that higher CS must universally imply better performance, but that comparing different forms of CS helps identify how coordination is organized in a given adaptive system.

As a control analysis, we also examined ordinary Pearson correlations computed directly from the raw adaptive threshold signals. Supplementary Fig.~\ref{fig:supp_rawcorr} shows the pair-specific correlations for all 15 threshold pairs as a function of the sensing-radius parameter $R$, together with the cooperation rate $C_r$. This analysis separates conventional linear similarity among the raw variables from synchronization of the time-dependent scaling dynamics used to define CS. While some threshold pairs exhibit substantial raw correlations, these correlations are highly pair-specific and do not provide a distributed description of coordination comparable to the CS network.

\FloatBarrier

\subsection*{Renewal-related supporting analyses}

\begin{figure*}[tbp]
\centering
\includegraphics[width=0.8\textwidth]{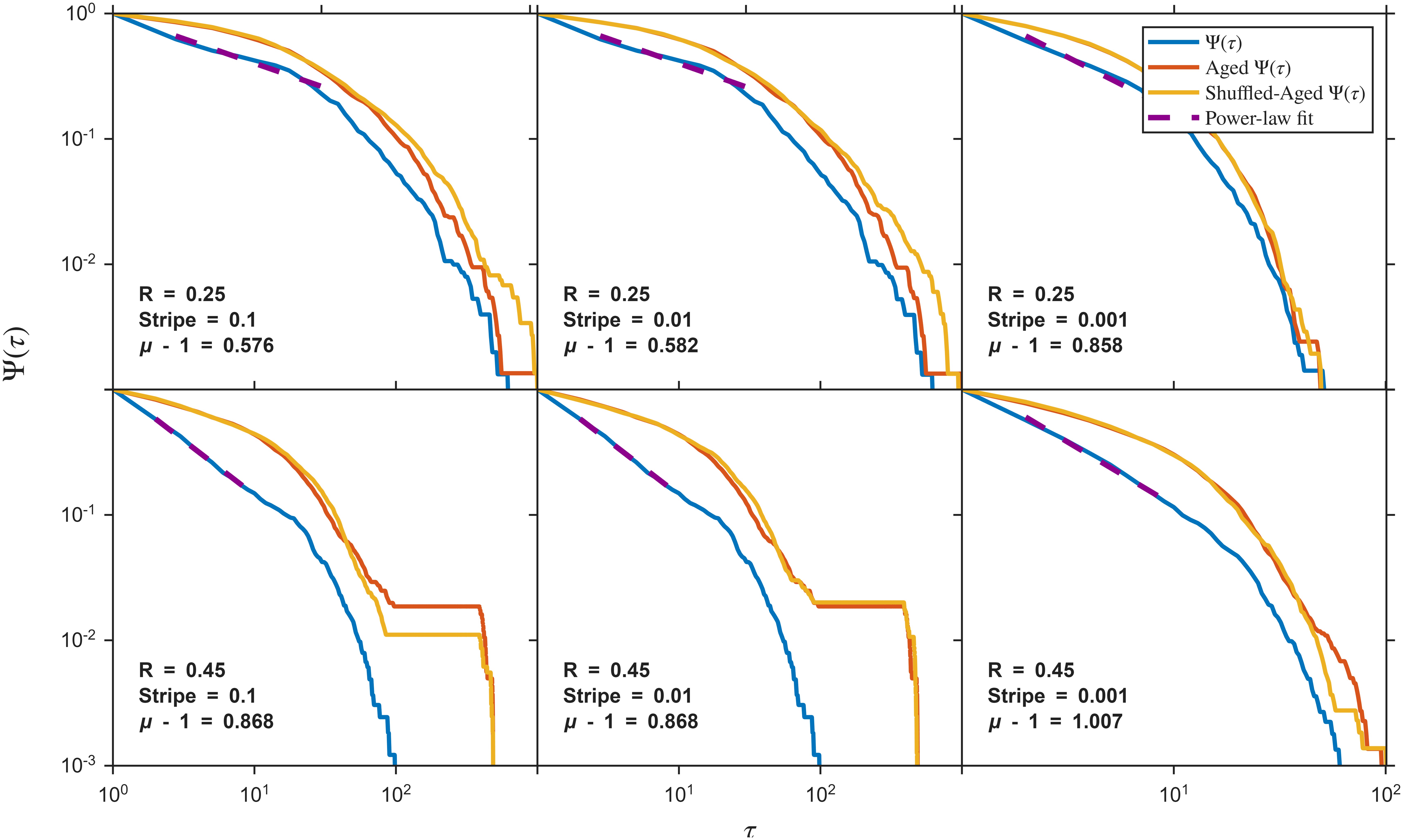}
\caption{Renewal aging test for the representative threshold $I_1$. Survival functions $\Psi(\tau)$, aged survival functions, and shuffled-aged controls are shown for two sensing-radius regimes ($R=0.25$ and $R=0.45$) and three stripe sizes. The first half of the simulation was discarded, and one post-discard window of length $10^4$ samples was analyzed. Events were defined as crossings from one stripe to another, and $\tau$ denotes the inter-event interval between successive stripe-crossing events. The aging time was fixed at $t_a=100$. The dashed black line shows the power-law fit used to estimate $\mu-1$.}
\label{fig:supp_renewal_test}
\end{figure*}

\begin{figure*}[tbp]
\centering
\includegraphics[width=0.8\textwidth]{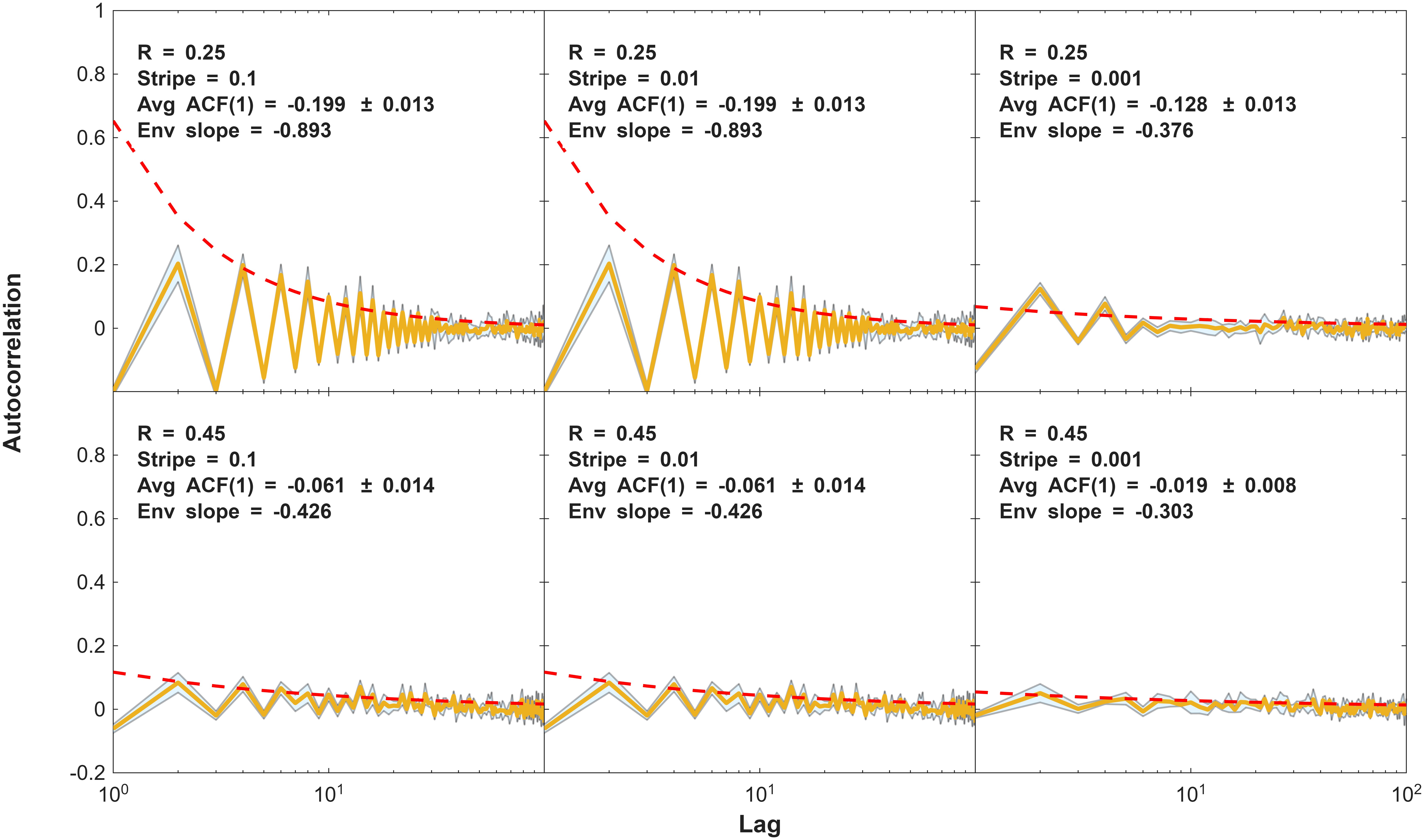}
\caption{Autocorrelation of inter-event intervals extracted from the representative threshold $I_1$. The first half of the simulation was discarded, and $\tau$ sequences were extracted from 10 windows of length $10^4$ samples with 75\% overlap for each sensing-radius and stripe-size condition. Autocorrelation was computed on the inter-event intervals $\tau$, not on the raw threshold signal. Curves show the mean across windows, and shaded bands indicate $\pm$ SD. The red dashed line shows a power-law fit to the upper envelope of the absolute autocorrelation, and the reported envelope slope summarizes the decay of inter-event memory with lag.}
\label{fig:supp_autotau}
\end{figure*}

To further interpret the temporal mechanism underlying the observed complexity-synchronization patterns, we examined whether stripe-crossing events extracted from the adaptive threshold dynamics exhibit renewal-like statistics. These analyses are used as supporting mechanistic evidence for the MDEA interpretation, rather than as additional definitions of CS. Specifically, Figs.~\ref{fig:supp_renewal_test} and~\ref{fig:supp_autotau} analyze the information-sharing threshold of Predator 1, $I_1$, as a representative adaptive threshold.

For the renewal aging test (Fig.~\ref{fig:supp_renewal_test}), events were extracted from $I_1$ using the same stripe-crossing method employed in MDEA. The first half of the simulation was discarded, and one post-discard window of length $10^4$ samples was analyzed. The inter-event interval $\tau$ denotes the number of samples between successive crossings from one stripe to another. Survival functions $\Psi(\tau)$ were compared with aged survival functions computed using aging time $t_a=100$ and with shuffled-aged controls. Because shuffling preserves the distribution of inter-event intervals while removing their temporal ordering, agreement between the aged and shuffled-aged curves suggests that the ordering of intervals contributes little additional memory, consistent with more renewal-like behavior.

In both sensing regimes ($R = 0.25$ and $R = 0.45$), the survival functions display scale-free behavior, consistent with heavy-tailed waiting-time statistics. The high-cooperation regime ($R = 0.45$) tends to exhibit closer agreement between the aged and shuffled-aged survival functions, particularly at finer stripe resolution, suggesting a reduction of temporal correlations and a closer approximation to renewal-like dynamics. In contrast, clearer discrepancies persist in the low-cooperation regime ($R = 0.25$), reflecting stronger inter-event memory.

Complementary evidence is provided by the autocorrelation of inter-event intervals (Fig.~\ref{fig:supp_autotau}). This analysis was again performed on $I_1$ after discarding the first half of the simulation, but used 10 windows of length $10^4$ samples with 75\% overlap. Autocorrelation was computed on the resulting $\tau$ sequences, not on the raw threshold signal. The plotted curves show the window-averaged autocorrelation, and the shaded bands show the standard deviation across windows. The red dashed line in each panel is a power-law fit to the upper envelope of the absolute autocorrelation; this envelope fit is used as a compact summary because the autocorrelation can oscillate around zero. The reported envelope slope summarizes the rate at which inter-event memory decays with lag. More negative slopes indicate faster decay.

In the low-cooperation regime, the inter-event interval series exhibits stronger oscillatory structure and a more slowly decaying envelope, suggesting persistent temporal correlations. By contrast, in the high-cooperation regime the autocorrelation rapidly decays toward zero beyond the first lag, consistent with effectively weaker memory between successive intervals.

Together, these results are consistent with the transition to higher cooperation being accompanied by a shift toward more renewal-like temporal organization in the representative $I_1$ threshold. This behavior supports the interpretation that MDEA-based CS captures coordination of event-driven restructuring dynamics, consistent with the positive relationship between CS and cooperation observed in Fig.~\ref{fig:cc_cs_mdea}.

\subsection*{CS as a diagnostic and control field}

The pairwise structure of CS provides more than a summary measure of coordination; it defines a diagnostic field over the interacting components of the system. Each pairwise CS value can be interpreted as the strength of alignment between the evolving scaling dynamics of two adaptive variables. In this view, CS is not a single global quantity, but a distributed field that reveals how coordination is organized across information sharing, trust, and payoff sharing.

\begin{figure*}[tbp]
\centering
\includegraphics[width=0.95\textwidth]{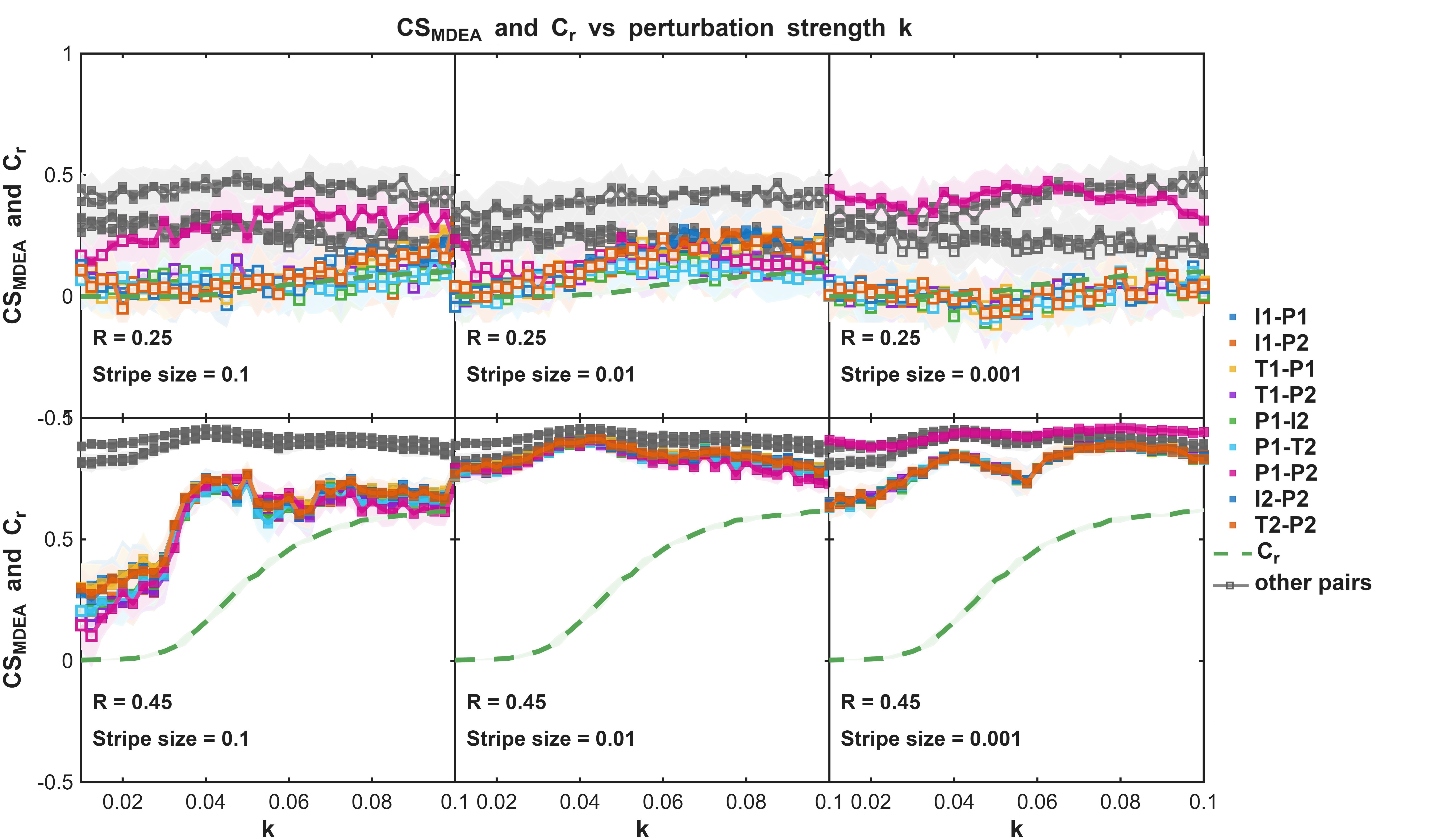}
\caption{
CS, computed using modified diffusion entropy analysis (MDEA), and cooperation rate ($C_r$) as functions of learning strength $k$ of the SharePay thresholds. The top row corresponds to $R = 0.25$ and the bottom row to $R = 0.45$. Columns represent stripe sizes of $0.1$, $0.01$, and $0.001$, respectively. Colored curves denote CS between selected threshold pairs $(I, T, P)$ involving the payoff sharing thresholds, while gray curves represent other pairwise combinations. The dashed green curve indicates the corresponding $C_r$. Shaded regions denote variability across ensembles.
}
\label{fig:CS_MDEA_k}
\end{figure*}

\begin{figure}[tbp]
\centering
\includegraphics[width=0.95\textwidth]{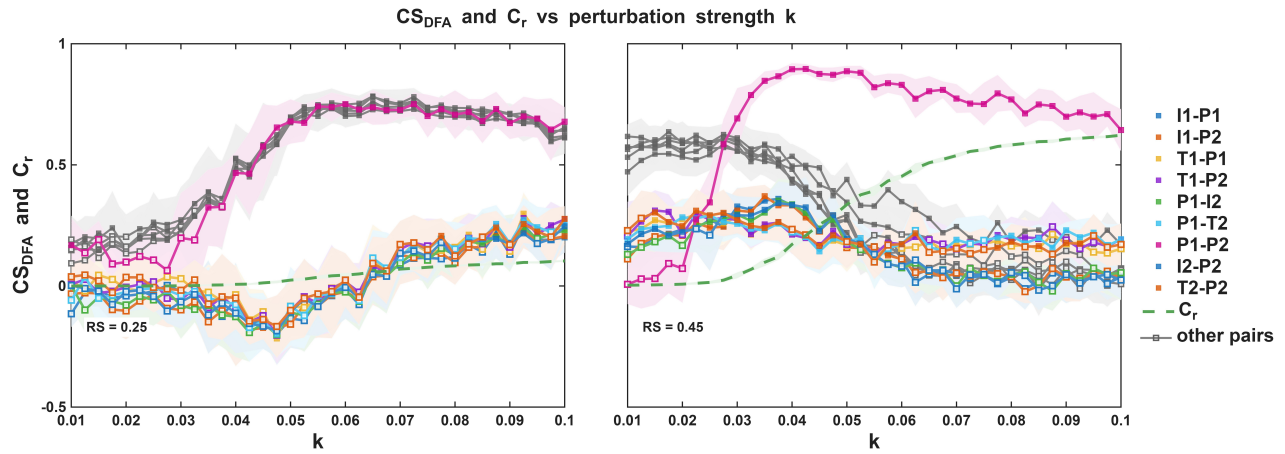}
\caption{
CS, computed using detrended fluctuation analysis (DFA), and  cooperation rate ($C_r$) as functions of learning strength $k$ of the SharePay thresholds. Panel (a) corresponds to $R = 0.25$ and panel (b) to $R = 0.45$. Colored curves represent CS between selected threshold pairs involving the payoff sharing thresholds, while gray curves denote other pairwise combinations. The dashed green curve indicates $C_r$. Shaded regions denote variability across ensembles.
}
\label{fig:CS_DFA_k}
\end{figure}

\paragraph{CS tracks regime-dependent changes in cooperation.}
We first examined how CS varies with the learning strength $k$ of the payoff-sharing thresholds across different sensing regimes. Figure~\ref{fig:CS_MDEA_k} shows that MDEA-based CS exhibits a clear dependence on $k$, with a strong contrast between low- and high-cooperation regimes. For $R = 0.25$, CS remains weak across threshold pairs and varies only modestly with $k$. For $R = 0.45$, CS increases sharply and organizes into a high-synchronization regime across multiple threshold pairs. The corresponding  cooperation rate $C_r$ increases with $k$ and reaches substantially higher values in the high-$R$ condition, co-occurring with the emergence of strong event-based CS.

The same analysis performed using DFA-based scaling (Fig.~\ref{fig:CS_DFA_k}) captures changes in coordination, but with weaker resolution in the high-cooperation regime. DFA-based CS remains more variable and less sharply organized than MDEA-based CS, indicating that the event-based scaling measure provides a more resolved description of the cooperative regime in this model. This contrast is consistent with the preceding results showing that cooperation is more strongly aligned with synchronized event-driven restructuring than with synchronized persistence alone.

\begin{figure*}[tbp]
\centering
\includegraphics[width=0.9\textwidth]{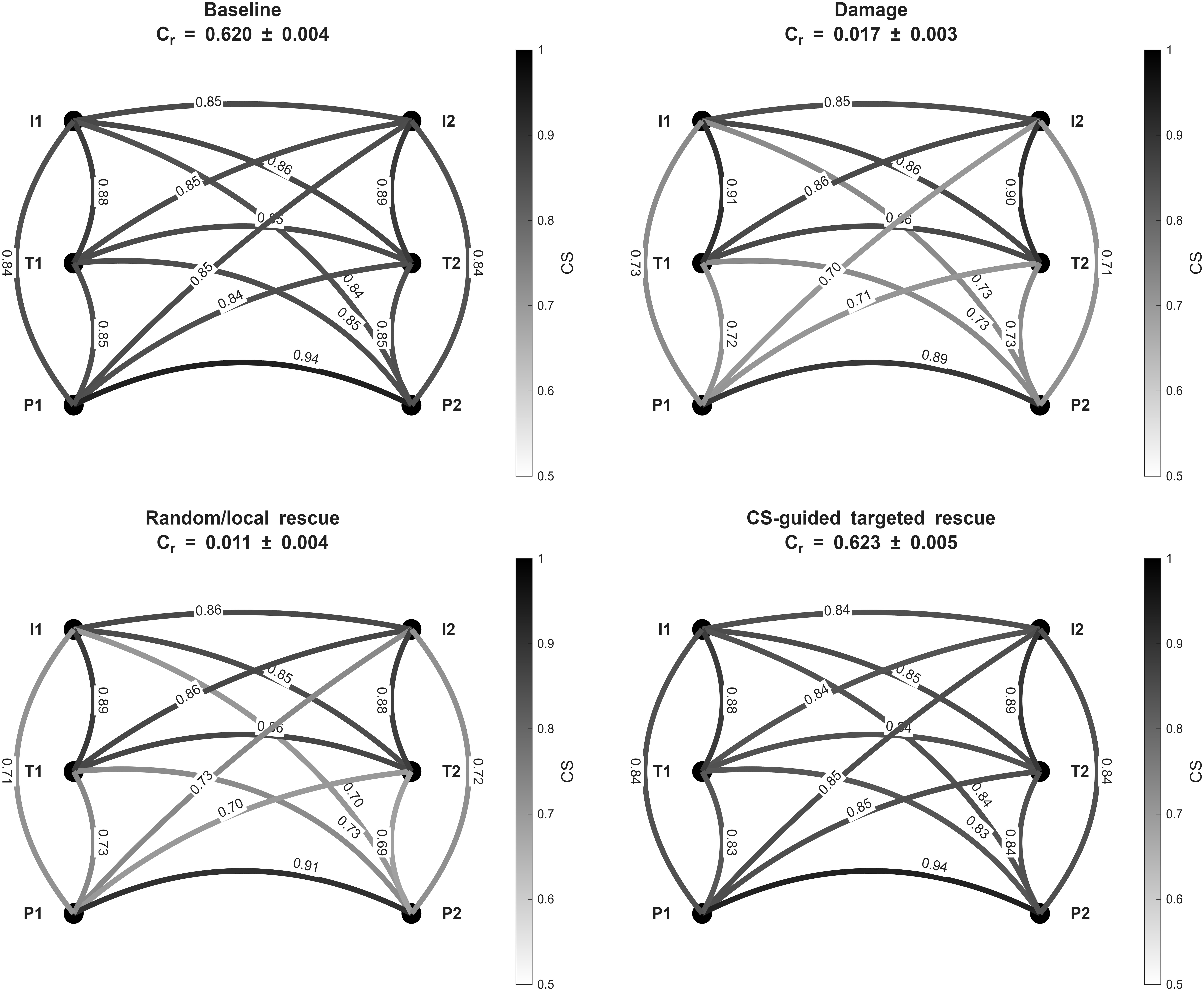}
\caption{
Complexity-synchronization (CS) networks under baseline, perturbation, and rescue conditions (MDEA, $R = 0.45$, ENS = 10). Graph representations of pairwise CS among the six adaptive thresholds $(I_1, T_1, P_1, I_2, T_2, P_2)$, computed using modified diffusion entropy analysis (MDEA). Nodes correspond to threshold variables, and edges represent CS links. Edge thickness and grayscale intensity encode CS strength, dashed edges denote non-significant links ($p \geq 0.05$), solid edges denote significant links ($p < 0.05$), and edge labels report the corresponding CS values. Panels show (top left) baseline, (top right) perturbed (damage), (bottom left) random/local rescue, and (bottom right) CS-guided targeted rescue. The  cooperation rate, $C_r$ (mean $\pm$ SD across ensembles), is reported in each panel. The grayscale color bar spans CS values from 0.5 to 1.
}
\label{fig:cs_rescue_network}
\end{figure*}

\begin{figure*}[tbp]
\centering
\includegraphics[width=0.8\textwidth]{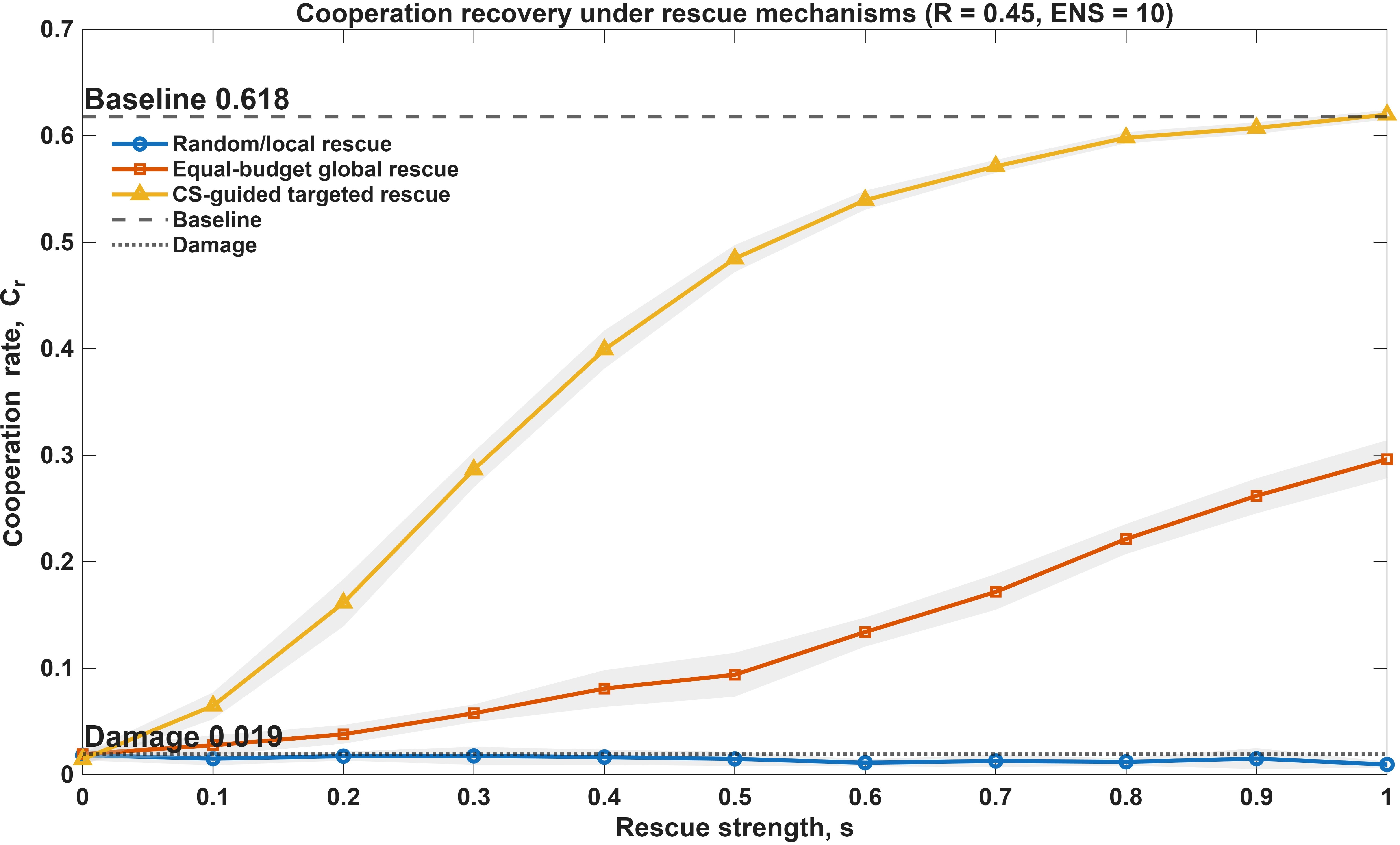}
\caption{
Cooperation recovery under different rescue strategies ($R = 0.45$, ENS = 10). Cooperation rate, $C_r$, is shown as a function of rescue strength, $s$, for three intervention strategies: random/local rescue, equal-budget global rescue, and CS-guided targeted intervention. Curves show the mean across ensembles, and shaded bands indicate $\pm$ SD. Horizontal dashed lines denote the baseline and perturbed (damage) levels.
}
\label{fig:rescue_curves}
\end{figure*}

\begin{figure*}[tbp]
\centering
\includegraphics[width=0.8\textwidth]{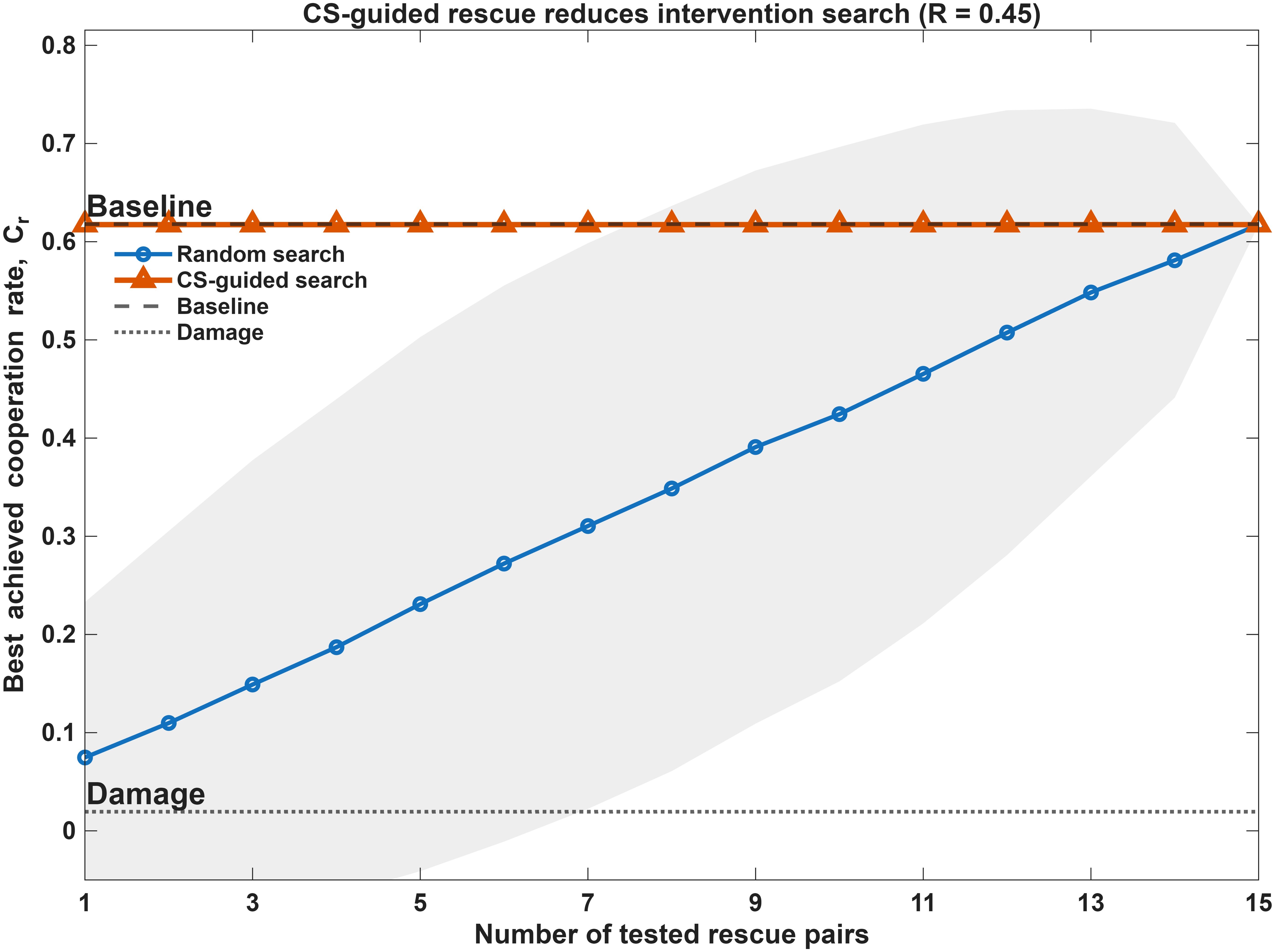}
\caption{
Search efficiency of CS-guided and random intervention strategies ($R = 0.45$). The best achieved cooperation rate, $C_r$, is plotted as a function of the number of tested threshold pairs (out of 15 possible pairs). The CS-guided strategy evaluates threshold pairs ranked according to the strength of their baseline complexity-synchronization (CS) links, whereas the random strategy samples threshold pairs uniformly. Curves show the mean across realizations, and the shaded region indicates $\pm$ SD. Horizontal dashed lines denote the baseline and perturbed (damage) levels.
}
\label{fig:search_efficiency}
\end{figure*}

\paragraph{CS localizes the functional effect of perturbation.}
We next asked whether the CS field can be used not only to describe cooperative organization, but also to guide intervention. We damaged the system by reducing the learning increment of the payoff-sharing thresholds $P_1$ and $P_2$ from $0.1$ to $0.025$, while leaving the other adaptive thresholds unchanged. At $R=0.45$, this perturbation reduced cooperation from the baseline level to a near-zero damaged state (Fig.~\ref{fig:cs_rescue_network}). The CS network also changed, but the change was not a uniform collapse of all links. Instead, the damaged condition retained substantial synchronization while showing weakened links involving the payoff-sharing subsystem. This dissociation indicates that CS and cooperation measure different aspects of the dynamics: CS describes the organization of adaptive temporal structure, whereas $C_r$ measures the behavioral function produced by that organization.

\paragraph{CS-guided targeted rescue restores cooperation.}
To test whether this diagnostic field is actionable, we compared three rescue strategies using the same damaged system. Random/local rescue applied the intervention budget to a non-targeted pair, equal-budget global rescue distributed the budget across all thresholds, and CS-guided targeted rescue restored the payoff-sharing pair identified from the CS structure. Figure~\ref{fig:rescue_curves} shows that targeted rescue recovered cooperation monotonically toward the baseline level as rescue strength increased. In contrast, random/local rescue remained close to the damaged level, while equal-budget global rescue produced only partial recovery. Thus, restoration of the specific payoff-sharing subsystem was sufficient to recover cooperative performance, whereas non-specific interventions were inefficient under the same task conditions.

\paragraph{CS reduces the intervention search space.}
Finally, we quantified whether CS-guided repair reduces the number of intervention tests required to restore cooperation. Figure~\ref{fig:search_efficiency} compares a CS-guided search, in which threshold pairs are evaluated according to their position in the CS structure, with a random search over pairs. For the perturbation considered here, for the perturbation considered here, the CS-guided strategy reached near-baseline cooperation with the first pair ranked by the CS-guided search, whereas random search improved only gradually as more pairs were tested and exhibited substantial variability across realizations. This result shows that the CS network reduces a combinatorial intervention problem to a low-dimensional control target. In this sense, CS provides not only a diagnostic description of cooperative organization, but also a practical guide for targeted repair.

The same intervention principle was examined in the low-$R$ regime, where cooperation is weaker and DFA-based synchronization provides greater contrast for the relevant coordination structure. The corresponding DFA-based rescue networks and recovery curves are reported in Supplementary Figs.~\ref{fig:supp_lowR_dfa_network} and~\ref{fig:supp_lowR_recovery}. Together, the main and supplementary rescue analyses show that CS can guide intervention across distinct coordination regimes, with MDEA providing the clearest diagnostic in the high-cooperation event-driven regime and DFA providing useful information in the low-$R$ regime.

Supporting renewal-related analyses are reported in Figs.~\ref{fig:supp_renewal_test} and~\ref{fig:supp_autotau}; their interpretation is discussed below as supporting mechanistic evidence rather than as the primary basis of the paper's central claim.

\FloatBarrier

\section*{Discussion and Implications for Complexity-Guided Team Control}\label{Discussion}

The results provide a unified picture of CS as both a diagnostic and an intervention-guiding principle for adaptive teams. The first part of the analysis showed that cooperation is associated with the synchronization of evolving scaling dynamics. In the present reduced Predator--Prey model, MDEA-based CS increases with cooperative performance, whereas DFA-based CS captures a different, persistence-related component of the dynamics. The second part of the analysis extends this diagnostic result by showing that the CS field can identify where intervention should be applied when cooperation fails.

The ordinary Pearson-correlation control analysis clarifies what is and is not captured by CS. Ordinary correlations quantify direct linear similarity between the raw adaptive thresholds, whereas CS quantifies synchronization of the evolving statistical structure extracted from those thresholds. The raw-correlation analysis reveals a highly heterogeneous pattern across threshold pairs. In particular, the strongest correlations are concentrated in a small subset of pairs, most notably the payoff-sharing pair $P_1$--$P_2$, whereas many other pairs remain weakly correlated or near zero across the entire $R$ range. By contrast, CS$_{\mathrm{MDEA}}$ reflects a distributed organization of adaptive dynamics across the network rather than the strength of individual raw signal correlations.

The rescue experiments are important because they move CS beyond a descriptive performance correlate. Perturbing the payoff-sharing learning channel collapses cooperation, but does not simply erase all synchronization in the adaptive network. This indicates that the system can retain coordinated temporal structure while losing the functional mapping from that structure to cooperative behavior. CS therefore should not be interpreted as identical to cooperation. Rather, CS resolves the organization of adaptive dynamics, and this organization can reveal which subsystem is functionally relevant for repair.

In the high-interaction regime ($R=0.45$), MDEA-based CS identified the payoff-sharing subsystem as an effective control target. Restoring this subsystem recovered cooperation to baseline levels, whereas random/local rescue failed and equal-budget global rescue produced only partial recovery. This result supports an intervention-guiding interpretation of CS: the network of synchronized scaling dynamics can reduce a high-dimensional intervention problem to a small set of candidate adaptive variables. In engineering terms, CS converts a blind search over intervention targets into a structured search guided by the coordination architecture of the system. This intervention result is demonstrated here for the specific payoff-sharing perturbation tested, and broader validation across additional perturbation types remains an important direction for future work.

The contrast between MDEA and DFA remains central to the interpretation. MDEA is sensitive to event-driven restructuring and therefore provides the clearest diagnostic in the high-cooperation regime, where adaptive organization is associated with renewal-like events. DFA emphasizes persistence and long-range correlation, which can be informative in regimes where memory-like structure dominates. The supplementary low-$R$ rescue analysis shows that the same intervention principle can be applied in such a regime using DFA-based CS. This result should be interpreted as complementary to, rather than competing with, the MDEA-based analysis. Thus, the value of CS is not that one scaling method is universally optimal, but that comparing scaling modes helps identify the temporal organization of coordination and the appropriate diagnostic representation for intervention.

The renewal-support analyses in Figs.~\ref{fig:supp_renewal_test} and~\ref{fig:supp_autotau} should be interpreted in this context. They do not define CS and they are not meant to establish renewal statistics for every threshold pair. Instead, they provide representative mechanistic support, using the $I_1$ threshold, for the idea that the high-cooperation regime is accompanied by weaker inter-event memory and a tendency toward more renewal-like temporal organization. The aging comparison in Fig.~\ref{fig:supp_renewal_test} suggests that temporal ordering of inter-event intervals contributes less additional structure in the high-$R$ regime, while the autocorrelation analysis in Fig.~\ref{fig:supp_autotau} shows faster decay of inter-event memory.

Interestingly, for the finest stripe size ($\mathrm{Str}=0.001$) in the high-cooperation regime ($R=0.45$), the estimated waiting-time exponent approaches $\mu \approx 2$. Using the relation $\beta = 3-\mu$ for renewal processes with heavy-tailed waiting times \cite{aquino2011transmission}, this corresponds to $\beta \approx 1$, characteristic of $1/f$-like temporal organization. Although this observation is based on a representative renewal analysis rather than an ensemble estimate, it is nevertheless consistent with the tendency toward scale-free event dynamics in the high-cooperation regime.

These observations help explain why an event-based complexity measure such as MDEA provides a clearer diagnostic of cooperative organization in the high-interaction regime, whereas DFA remains useful for regimes in which persistence-dominated dynamics are more prominent.

This interpretation also clarifies the distinction between CS and complexity matching. In complexity matching, information transfer is favored when interacting systems have similar complexity levels \cite{aquino2011transmission,west2008}. In CS, by contrast, the changes in complexity become synchronized under interaction \cite{mahmoodi2023complexity,west2023complexity}. The present results support the latter mechanism and extend it: synchronized changes in complexity can be used not only to characterize coordination, but also to identify leverage points for restoring cooperative function.

More broadly, these findings suggest that adaptive intelligence is not simply a state of high correlation or high performance. It is a process in which interacting components reorganize their temporal complexity in a coordinated manner, and in which the functional consequences of that organization depend on specific adaptive channels. For human--machine collaboration, rehabilitation, and engineered collective systems, this distinction is important. Standard performance variables indicate whether a system succeeds, but not which part of the system should be modified when performance degrades. CS provides a candidate framework for identifying such targets.

The present demonstration is model-based, and the specific control target identified here---the payoff-sharing subsystem---is tied to the structure of the reduced Predator--Prey task. The broader principle is more general: when adaptive systems contain interacting internal variables, the synchronization of their evolving complexity can define a diagnostic field over possible intervention targets. Future work should test whether the same principle can guide repair in empirical human--human and human--machine teaming data, where the relevant variables may be neural, behavioral, physiological, or task-derived rather than explicit model thresholds.

\section*{Conclusion}\label{sec4}

This study demonstrates that CS provides a framework for diagnosing and repairing cooperative organization in adaptive systems. Using SA-based multi-agent dynamics in a reduced Predator--Prey model, we showed that cooperation is associated with synchronized scaling dynamics and that MDEA- and DFA-based CS reveal distinct temporal modes of coordination.

The central advance is that CS can be used as an actionable diagnostic field. When payoff-sharing learning was perturbed, cooperation collapsed. CS-guided targeted rescue restored cooperation to baseline levels, whereas random/local rescue failed and equal-budget global rescue produced only partial recovery. Thus, CS does more than quantify coordination: it identifies adaptive subsystems that can be targeted to recover function.

These findings suggest that CS may serve as a diagnostic and engineering principle for adaptive teams. By revealing how coordination is organized across interacting components, CS provides a potential route from measurement to intervention in biological, social, human--machine, and engineered collective systems.

\clearpage
\section*{Supplementary Information}

The following supplementary sections retain the detailed model description and implementation details so that the main paper can remain focused on the conceptual result.

\subsection*{S1. Detailed Predator--Prey model and simulation parameters}

The simulations use one Prey agent and two Predator agents moving on a square domain of length $L=2$ with periodic boundary conditions. The discrete time step is $\Delta t = 1$. The Prey and Predators update their positions using constant-speed motion on the torus, with the Predators moving at speed $v_S=0.1$ and the Prey moving at speed $v_F=0.2$.

The dimensionless sensing-radius ratio is defined as
\[
R = \frac{r_S}{L},
\]
and is swept over
\[
R = 0.24 + 0.01(1:21),
\]
so that the effective Predator sensing radius is $r_S = R \times L$. The Prey uses a comparable sensing radius and chooses an evasive heading by predicting the mean Predator position and applying a deceptive angular deflection of $\pm \pi/6$.

Each Predator carries three adaptive thresholds associated with information sharing, trust, and payoff sharing. These thresholds are denoted by
\[
I_1,\, T_1,\, P_1,\, I_2,\, T_2,\, P_2.
\]
At each time step, binary decisions are generated by comparing each threshold to a uniform random number. The thresholds are then updated using payoff-based reinforcement, with learning step
\[
\Delta_S = 0.1
\]
and a small stochastic term with amplitude
\[
\mathrm{Noise} = 10^{-3}.
\]

When at least one Predator successfully captures the Prey and trust is active, the two Predators enter a payoff-sharing interaction governed by a Prisoner’s Dilemma–like payoff matrix. Writing the row player as Predator 1 and the column player as Predator 2, the payoff matrix is

\begin{equation}
\begin{array}{c|cc}
 & \text{Predator 2: Cooperate} & \text{Predator 2: Defect} \\
\hline
\text{Predator 1: Cooperate} & (2,\,2) & (-2,\,5) \\
\text{Predator 1: Defect}    & (5,\,-2) & (-1,\,-1)
\end{array}
\end{equation}

Cooperation corresponds to both agents share payoff $(2,2)$, and the  cooperation rate $C_r$ is defined as the cumulative sum of such events over time. Unilateral defection yields the temptation--sucker outcomes $(5,-2)$ and $(-2,5)$, and mutual defection yields the punishment pair $(-1,-1)$. In the simulations reported here, the temptation parameter is set by $tc=3$.

For each parameter condition, the simulation runs for
\[
T = 10^6
\]
trials and
\[
ENS = 10
\]
ensembles. After discarding the first 25\% of the threshold trajectories, local scaling exponents are computed on overlapping windows of length
\[
\mathrm{Slice}=10^4
\]
with 75\% overlap. MDEA is evaluated for the three representative stripe sizes
\[
\{0.1,\ 0.01,\ 0.001\}. 
\] These stripe sizes were selected to span coarse, intermediate, and fine event resolutions and were used consistently throughout all MDEA analyses.

Unless otherwise stated, MDEA was implemented using Rule = 1, and local scaling exponents were obtained from the entropy-growth curve within the fitting interval used throughout the simulations.

In the implementation used for the $C_r--CS$ figures, the reported pairwise CS quantities correspond to three representative threshold pairs extracted from the scaling-exponent correlation matrix. The  cooperation rate $C_r$ is computed as the running fraction of mutual payoff-sharing events per time.

\subsection*{S2. Explicit decision rules and update equations}

This section provides the explicit dynamical rules underlying the simulations, including spatial geometry, directional updates, stochastic decision-making, payoff assignment, and threshold adaptation.

\paragraph*{Periodic geometry and directional angles.}
All agents move on a square domain of side length $L$ with periodic boundary conditions. For two positions
\[
\mathbf{r}_a=(x_a,y_a), \qquad \mathbf{r}_b=(x_b,y_b),
\]
the minimal-image displacement is
\[
\Delta x_{ab}=x_b-x_a-L\,\mathrm{round}\!\left(\frac{x_b-x_a}{L}\right), \qquad
\Delta y_{ab}=y_b-y_a-L\,\mathrm{round}\!\left(\frac{y_b-y_a}{L}\right).
\]
The corresponding direction from $a$ to $b$ is
\[
\Theta(\mathbf{r}_a \rightarrow \mathbf{r}_b)=\operatorname{atan2}(\Delta y_{ab},\Delta x_{ab}),
\]
and the torus-consistent distance is
\[
d_{ab}=\sqrt{\Delta x_{ab}^2+\Delta y_{ab}^2}.
\]

\paragraph*{State variables and motion.}
The system consists of one Prey agent $F$ and two Predator agents $S_1$ and $S_2$. Their positions and headings evolve as
\[
\mathbf{r}(t+\Delta t)=\mathbf{r}(t)+v
\begin{bmatrix}
\cos\theta(t)\\
\sin\theta(t)
\end{bmatrix}\Delta t
\pmod L.
\]
Predators move with speed $v_S$ and the Prey with speed $v_F=2v_S$. $\Delta t=1$.

\paragraph*{Prey directional update and deception.}
The Prey senses Predators within its vision radius $r_F$. If at least one Predator is detected, it predicts their mean future position
\[
\widehat{\mathbf{r}}_{S}(t+\Delta t)
=
\frac{1}{N}\sum_{j}
\left(
\mathbf{r}_{S_j}(t)+
v_S
\begin{bmatrix}
\cos\theta_{S_j}(t)\\
\sin\theta_{S_j}(t)
\end{bmatrix}
\right),
\]
where the sum runs over detected Predators.

The Prey computes the direction toward this estimate:
\[
\phi_F(t)=\Theta(\mathbf{r}_F(t)\rightarrow \widehat{\mathbf{r}}_{S}(t+\Delta t)).
\]

It then generates two symmetric evasive headings:
\[
\theta_F^{(\pm)}(t)=\pi+\phi_F(t)\pm \Theta_D,
\]
where $\Theta_D=\pi/6$.

The realized heading is chosen stochastically:
\[
\theta_F(t)\in\{\theta_F^{(+)},\theta_F^{(-)}\}, \quad P=1/2.
\]

This two-option rule introduces deception and stochasticity in the environment.

\paragraph*{Predator directional prediction.}
If the Prey is within sensing radius $r_S$, each Predator $S_i$ predicts two possible future Prey positions:
\[
\widehat{\mathbf{r}}_{F}^{(\pm)}(t+\Delta t)
=
\mathbf{r}_F(t)+
v_F
\begin{bmatrix}
\cos\theta_F^{(\pm)}(t)\\
\sin\theta_F^{(\pm)}(t)
\end{bmatrix}.
\]

These define two candidate pursuit directions:
\[
\phi_{S_i}^{(\pm)}(t)=
\Theta(\mathbf{r}_{S_i}(t)\rightarrow \widehat{\mathbf{r}}_{F}^{(\pm)}(t+\Delta t)).
\]

Each Predator randomly selects one as its own estimate:
\[
\theta_{S_i}^{\mathrm{self}}(t)\in\{\phi_{S_i}^{(+)},\phi_{S_i}^{(-)}\},
\]
while the other is stored as the complementary estimate:
\[
\theta_{S_i}^{\mathrm{comp}}(t).
\]

\paragraph*{Probabilistic decision rules.}
Each Predator carries three thresholds:
\[
I_i(t), \quad T_i(t), \quad P_i(t).
\]

Each generates a binary decision via
\[
s_i^{(X)}(t)=
\begin{cases}
1, & u > X_i(t),\\
0, & \text{otherwise},
\end{cases}
\quad u\sim\mathcal{U}(0,1).
\]

Here:
\[
s_i^{(I)}=1 \rightarrow \text{share information}, \quad
s_i^{(T)}=1 \rightarrow \text{trust}, \quad
s_i^{(P)}=1 \rightarrow \text{share payoff}.
\]

\paragraph*{Trust-mediated complementary coupling.}
Let $j\neq i$. The realized heading of Predator $S_i$ is
\[
\theta_{S_i}(t)=
\begin{cases}
\theta_{S_j}^{\mathrm{comp}}(t), & s_i^{(T)}(t)=1,\\
\theta_{S_i}^{\mathrm{self}}(t), & s_i^{(T)}(t)=0.
\end{cases}
\]

Thus, when trust is active, a Predator adopts the complementary estimate from its partner, allowing the pair to cover both possible Prey trajectories.

\paragraph*{Environmental capture and stage-1 payoff.}
Capture is determined by proximity:
\[
\chi_i(t)=
\begin{cases}
1, & d(\mathbf{r}_{S_i},\mathbf{r}_F)<r_G,\\
0, & \text{otherwise}.
\end{cases}
\]

If $\chi_i(t)=1$, capture succeeds with probability $p_{\mathrm{env}}$. Otherwise, it fails.

The resulting environmental payoff is
\[
\Pi_i^{(0)}(t)=
\begin{cases}
+2, & \text{success},\\
-2, & \text{failure}.
\end{cases}
\]

\paragraph{Definition of  cooperation rate $C_r$.}

At each time step $t$, a payoff-sharing interaction between the two Predators is evaluated only under specific conditions.

First, a successful capture event must occur, defined by
\[
\chi_i(t) = 1 \quad \text{for at least one } i \in \{1,2\},
\]
indicating that at least one Predator captures the Prey.

Second, the interaction must be enabled by trust. In the implementation used here, this condition is satisfied if at least one of the two Predators realizes a trust decision,
\[
s^{(T)}_1(t) = 1 \quad \text{or} \quad s^{(T)}_2(t) = 1.
\]

When both conditions are satisfied, the system enters a payoff-sharing stage in which each Predator makes a binary decision regarding payoff sharing:
\[
s^{(P)}_i(t) =
\begin{cases}
1, & \text{share payoff}, \\
0, & \text{do not share}.
\end{cases}
\]

Mutual cooperation is defined as the simultaneous realization
\[
s^{(P)}_1(t) = s^{(P)}_2(t) = 1.
\]

The cooperation rate $C_r$ is then computed as the running fraction of such mutual-cooperation events over time,
\[
\mathrm{C_r}(t) = \frac{1}{t} \sum_{\tau=1}^{t}
\mathbb{I}\Big[
\text{capture}(\tau)
\;\wedge\;
\big(s^{(T)}_1(\tau) = 1 \;\vee\; s^{(T)}_2(\tau) = 1\big)
\;\wedge\;
\big(s^{(P)}_1(\tau) = 1 \;\wedge\; s^{(P)}_2(\tau) = 1\big)
\Big],
\]
where $\mathbb{I}[\cdot]$ denotes the indicator function. This definition reflects the fact that cooperation is evaluated only within
interaction events that are both environmentally successful and socially enabled.

\paragraph*{Stage-2 Prisoner’s Dilemma interaction.}
A payoff-sharing interaction occurs if at least one Predator succeeds and at least one trusts.

Given decisions $s_1^{(P)}, s_2^{(P)}$, the payoff matrix is
\[
(\Pi_1,\Pi_2)=
\begin{cases}
(2,2), & (1,1),\\
(2+t_c,-2), & (0,1),\\
(-2,2+t_c), & (1,0),\\
(-1,-1), & (0,0),
\end{cases}
\]
with $t_c=3$.

Otherwise, $\Pi_i(t)=\Pi_i^{(0)}(t)$.

\paragraph*{Threshold adaptation.}
Each threshold evolves via payoff-driven reinforcement:
\[
X_i(t+1)=\mathcal{P}_{[0,1]}\left[
X_i(t)+\sigma(s_i)\,\delta_X\,(\Delta\Pi_i+\eta)
\right],
\]
where
\[
\sigma(s)=
\begin{cases}
-1, & s=1,\\
+1, & s=0,
\end{cases}
\quad
\eta\sim\mathcal{N}(0,\text{Noise}^2).
\]

$\mathcal{P}_{[0,1]}$ projects values into $[0,1]$.

\paragraph*{Complexity-synchronization pipeline.}
The six threshold time series
\[
I_1,\ T_1,\ P_1,\ I_2,\ T_2,\ P_2
\]
are analyzed using sliding windows. Local scaling exponents $\delta(t)$ are computed via MDEA or DFA, and CS is defined as
\[
CS=\mathrm{corr}(\delta_a(t),\delta_b(t)).
\]

\subsection*{S3. Robustness of the intervention principle in the low-$R$ regime}

The main rescue analysis was performed at $R=0.45$, where cooperation is high and MDEA-based CS provides the most resolved description of the event-driven coordination structure. To test whether the intervention principle also applies in a qualitatively different regime, we repeated the rescue analysis at $R=0.25$. In this low-interaction regime, cooperation is weaker and MDEA-based CS is less pronounced. DFA-based CS provides greater contrast because the dynamics contain a stronger persistence-related component.

The same perturbation and rescue logic was used as in the main text. Damage was imposed by reducing the learning increments of the payoff-sharing thresholds $P_1$ and $P_2$. Random/local rescue applied the intervention budget to a non-targeted pair, equal-budget global rescue distributed the budget across thresholds, and CS-guided targeted rescue restored the payoff-sharing subsystem identified by the CS structure. All simulations used $10^6$ trials and ENS = 10 independent stochastic ensembles.

\begin{figure*}[tbp]
\centering
\includegraphics[width=1.1\textwidth]{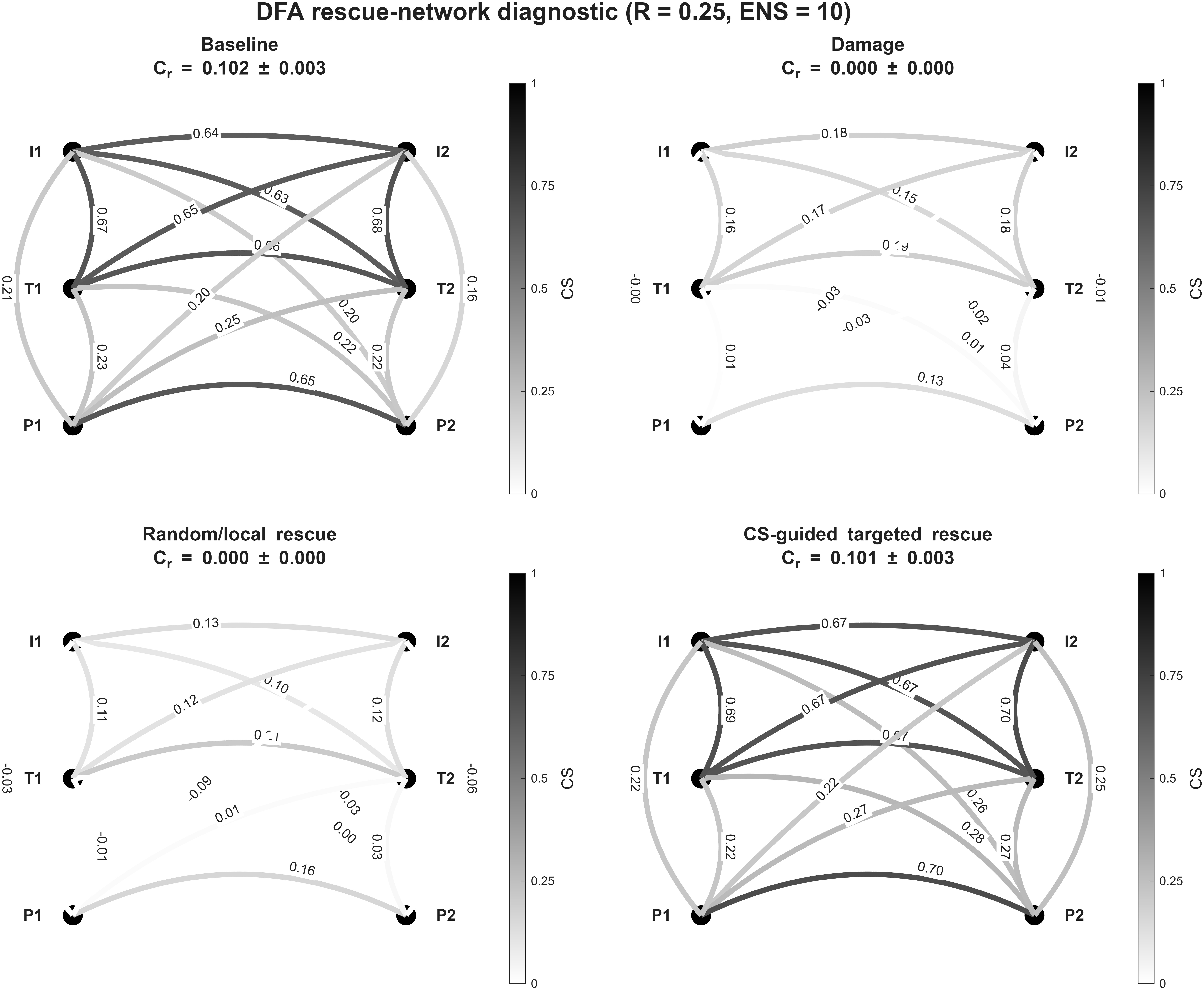}
\caption{\textbf{Supplementary Fig. S4.} DFA-based complexity-synchronization (CS) networks under baseline, perturbation, and rescue conditions ($R = 0.25$, ENS = 10). Graph representations of pairwise CS among the six adaptive thresholds $(I_1, T_1, P_1, I_2, T_2, P_2)$, computed using detrended fluctuation analysis (DFA). Nodes correspond to threshold variables, and edges represent CS links. Edge thickness and grayscale intensity encode CS strength, dashed edges denote non-significant links ($p \geq 0.05$), solid edges denote significant links ($p < 0.05$), and edge labels report the corresponding CS values. Panels show (top left) baseline, (top right) perturbed (damage), (bottom left) random/local rescue, and (bottom right) CS-guided targeted rescue. The cooperation rate, $C_r$ (mean $\pm$ SD across ensembles), is reported in each panel. The color scale spans the interval $[0,1]$.}
\label{fig:supp_lowR_dfa_network}
\end{figure*}

\begin{figure*}[tbp]
\centering
\includegraphics[width=0.8\textwidth]{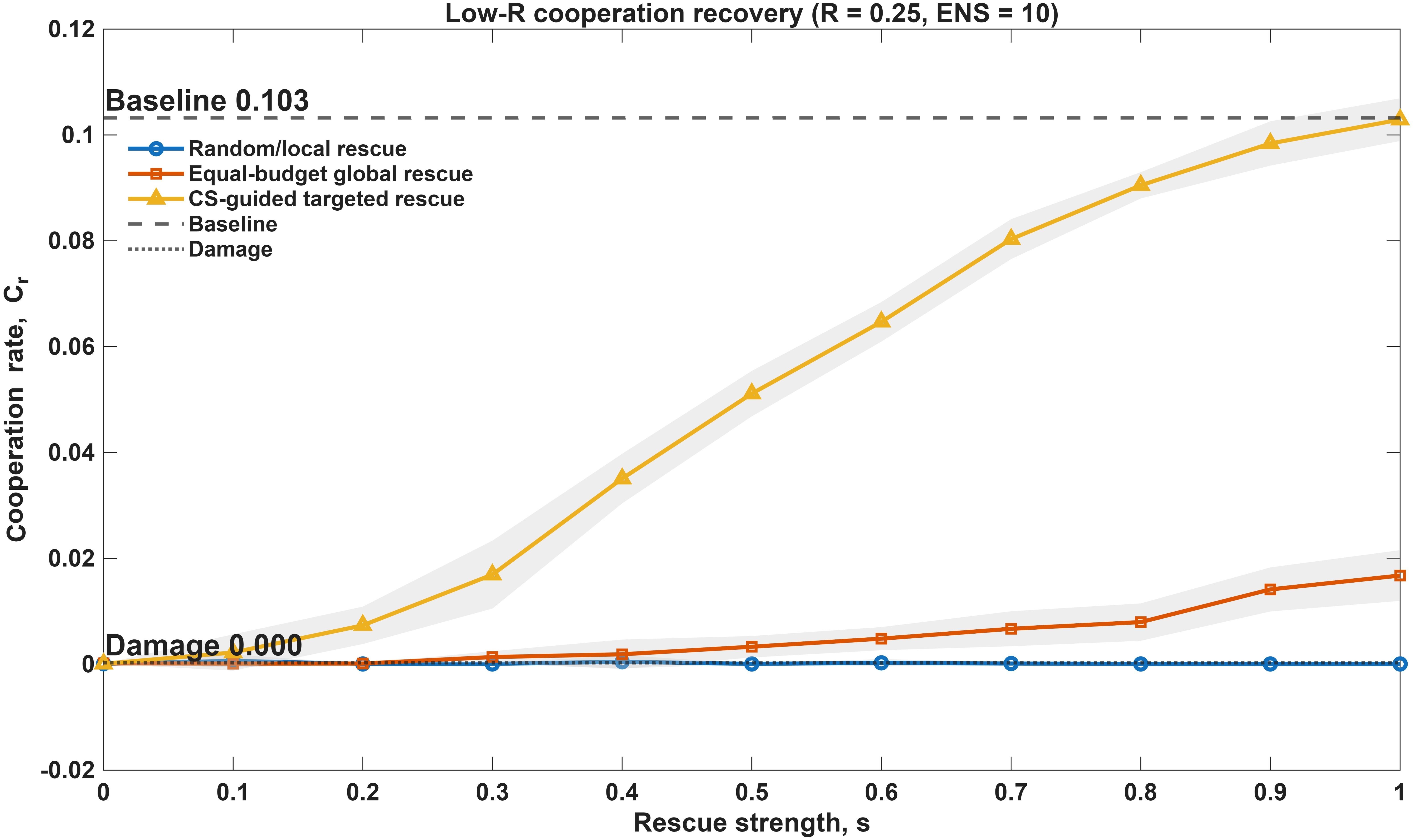}
\caption{\textbf{Supplementary Fig. S5.} Cooperation recovery in the low-interaction regime ($R = 0.25$, ENS = 10) using DFA-based complexity-synchronization (CS). The  cooperation rate, $C_r$, is shown as a function of rescue strength, $s$, for three intervention strategies: random/local rescue, equal-budget global rescue, and CS-guided targeted intervention. Curves show the mean across ensembles, and shaded bands indicate $\pm$ SD. Horizontal dashed lines denote the baseline and perturbed (damage) levels.}
\label{fig:supp_lowR_recovery}
\end{figure*}

Supplementary Fig.~\ref{fig:supp_lowR_dfa_network} shows that DFA-based CS is weak and heterogeneous in the damaged condition and in the random/local rescue condition, whereas the targeted rescue condition recovers a network structure closer to the baseline low-$R$ state. Supplementary Fig.~\ref{fig:supp_lowR_recovery} shows the corresponding cooperation recovery curves. Targeted rescue increases $C_r$ toward the baseline level as rescue strength increases, while random/local rescue remains near the damaged level and equal-budget global rescue produces only a small improvement. These results support the interpretation that the CS-guided intervention principle is not restricted to the high-$R$ MDEA regime. Instead, the appropriate scaling representation depends on the coordination mode of the system: MDEA is most informative for event-driven cooperative organization, whereas DFA can provide useful diagnostic information in the low-interaction regime.

\subsection*{S4. Modified Diffusion Entropy Analysis (MDEA)}

\paragraph{Diffusion construction from event sequences}

To quantify temporal complexity, we employ Modified Diffusion Entropy Analysis (MDEA), which converts each time series into a diffusion process driven by discrete events. Events are extracted using a stripe-based method, whereby crossings of predefined amplitude levels generate a sequence of event times. These events are then mapped onto a diffusion trajectory by assigning unit increments at each event occurrence, resulting in an effective random walk.

\paragraph{Scaling hypothesis}

The diffusion process is assumed to satisfy the scaling form:
\begin{equation}
F(x,t) = \frac{1}{t^{\delta}} \, \Phi\!\left(\frac{x}{t^{\delta}}\right),
\end{equation}
where $F(x,t)$ denotes the probability density function (PDF) of the diffusion variable $x$ at time $t$, $\delta$ is the scaling exponent, and $\Phi(\cdot)$ is a time-independent scaling function. This form implies statistical self-similarity, meaning that the shape of the distribution remains invariant under appropriate rescaling. Consequently, the characteristic spread of the process follows $x \sim t^{\delta}$.

\paragraph{Entropy scaling}

The scaling exponent is obtained through the Shannon entropy of the diffusion process:
\begin{equation}
S(t) = -\int F(x,t)\,\log F(x,t)\,dx.
\end{equation}
Substituting the scaling form of $F(x,t)$ leads to:
\begin{equation}
S(t) = A + \delta \log t,
\end{equation}
where $A$ is a constant. The exponent $\delta$ is estimated as the slope of $S(t)$ versus $\log t$. This entropy-based approach remains valid for non-Gaussian processes and is particularly effective for signals driven by intermittent or heavy-tailed events.

\paragraph{Relation to renewal dynamics}

For renewal processes characterized by inverse power-law waiting-time distributions,
\begin{equation}
\psi(\tau) \sim \tau^{-\mu}, \quad 2 < \mu < 3,
\end{equation}
the scaling exponent is related to $\mu$ by:
\begin{equation}
\delta = \frac{1}{\mu - 1}.
\end{equation}
This relationship links the diffusion scaling directly to the temporal statistics of events, enabling identification of long-range temporal organization.

\paragraph{Application in this study}

MDEA is applied within sliding windows to obtain a time-resolved scaling exponent $\delta(t)$ for each signal. These scaling time series are interpreted as measures of instantaneous complexity. CS is then quantified as the statistical association between the scaling exponents of interacting components.

\subsection*{S5. Detrended Fluctuation Analysis (DFA)}

\paragraph{Profile construction}

Detrended Fluctuation Analysis (DFA) is used as an alternative method to quantify scaling behavior. Given a time series $x(i)$ of length $N$, the first step is to construct the integrated profile:
\begin{equation}
Y(k) = \sum_{i=1}^{k} \left(x(i) - \langle x \rangle \right),
\end{equation}
where $\langle x \rangle$ is the mean of the signal.

\paragraph{Segmentation and detrending}

The profile $Y(k)$ is divided into non-overlapping segments of equal length $n$. Within each segment, a polynomial trend (typically linear in this study) is fitted and removed. The root-mean-square fluctuation is then computed as:
\begin{equation}
F(n) = \sqrt{\frac{1}{N} \sum_{k=1}^{N} \left[Y(k) - Y_n(k)\right]^2},
\end{equation}
where $Y_n(k)$ denotes the local trend in each segment.

\paragraph{Scaling relationship}

If the signal exhibits scale-invariant behavior, the fluctuation function follows:
\begin{equation}
F(n) \sim n^{\alpha},
\end{equation}
where $H$ is the DFA scaling exponent. The exponent is obtained as the slope of $\log F(n)$ versus $\log n$.

\paragraph{Interpretation}

The value of $H$ characterizes the correlation structure of the signal:
\begin{itemize}
    \item $\alpha = 0.5$: uncorrelated (white noise)
    \item $\alpha > 0.5$: persistent correlations
    \item $\alpha < 0.5$: anti-persistent behavior
\end{itemize}

\paragraph{Application in this study}

DFA is applied within sliding windows to obtain a time-resolved scaling exponent $H(t)$ for each signal. Similar to MDEA, these scaling time series are used to quantify complexity. CS is computed as the statistical association between the DFA scaling exponents of interacting components.

\paragraph{Complementarity with MDEA}

While DFA captures correlation-based scaling properties of the signal amplitude, MDEA characterizes the scaling behavior of event-driven diffusion processes. The two approaches therefore probe distinct aspects of temporal organization, and their combined use provides a more comprehensive description of system dynamics.

It is important to stress that DFA is not a proper method to evaluate scaling, but should be interpreted as a method to evaluate the second moment of a distribution density whose scaling can be properly evaluated using MDEA \cite{kalashyan2009ergodicity}. The adoption of the arguments illustrated in this paper lead to

\begin{equation}
    H = \frac{4 - \mu}{2},
\end{equation}
which for $\mu < 2$ yields $H > 1$. 

It is important to notice that \cite{kalashyan2009ergodicity} is based on the adoption of the velocity model \cite{giacomoDEA2} , filling the time region between two consecutive crucial events with either $W$ or $-w$, where $W > 0$. The DEA analysis adopted in this paper is done using DEA with stripes \cite{allegrini2002memory}.It is important to notice that the time region between two consecutive crucial events, called laminar region, is not empty, but filled by non-crucial events, that may either totally random or correlated without hosting crucial events. The adoption of DEA, with the de-trending of the non-crucial events hosted by the laminar region leads to Eq.(12),
which for $\mu < 2$ yields $H > 1$. 

In the case where the laminar regions do not host any event the prescription of \cite{giacomoDEA2} yields as scaling the result
\begin{equation}
    \delta  = \mu -1,
\end{equation}
which is smaller than $1$ if $\mu < 2$.

There is an important connection with quantum mechanics \cite{Sabin2026} that suggests further research work to do establish a stronger connection between the results of this paper and the open issue of cognition. The work \cite{Sabin2026} shows that the prediction of Eq. (12) is very close to 
In the case where the laminar regions do not host any event the prescription of \cite{giacomoDEA2} yields as scaling the result
\begin{equation}
    \delta  = \frac{1}{\mu-1},
\end{equation}
when the condition $2 < \mu < 3$ applies. In this case no strong conflict exists between $DEA$ and $DFA$. However, the slight difference between Eq. (12) and Eq. (14) has been satisfactorily explained by the work of \cite{density1} and \cite{density2}, as a conflict between the analysis of stochastic trajectories and the corresponding density approach. The divergence between DEA and DFA becomes much stronger when $\mu < 2$.

\subsection*{S6. Ordinary Pearson-correlation control analysis}

To compare complexity synchronization with conventional linear similarity, ordinary Pearson correlations were computed directly from the raw adaptive threshold signals after the initial burn-in period. Pairwise Pearson correlations were evaluated for all 15 threshold pairs among the six adaptive variables ($I_1$, $I_2$, $T_1$, $T_2$, $P_1$, and $P_2$), and the ensemble-averaged pair-specific correlations were examined as functions of the sensing-radius parameter $R$. This analysis asks whether direct amplitude-level correlations among the adaptive thresholds reproduce the coordination structure revealed by CS. The results are summarized in Supplementary Fig.~\ref{fig:supp_rawcorr}.

\begin{figure*}[tbp]
\centering
\includegraphics[width=0.95\textwidth]{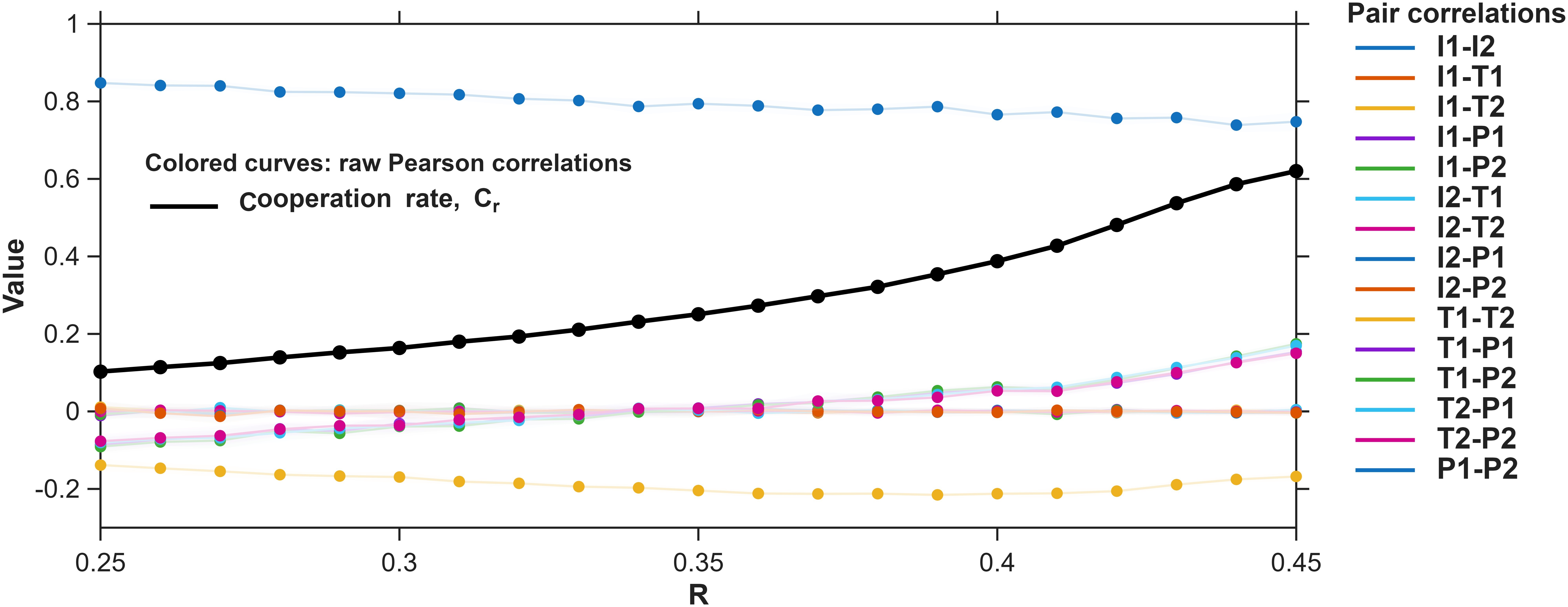}
\caption{\textbf{Supplementary Fig. S6. Ordinary Pearson correlations as a function of sensory radius.}
The ensemble-averaged  cooperation rate ($C_r$, black) and ordinary Pearson correlations (colored curves) are shown as functions of the sensory-radius parameter $R$. Colored curves represent the ordinary Pearson correlation computed for each of the 15 unique pairs among the six adaptive variables ($I_1$, $I_2$, $T_1$, $T_2$, $P_1$, and $P_2$), and the right-hand legend identifies the corresponding variable pair for each curve. The figure shows that raw correlations are highly pair-specific: the payoff-sharing pair $P_1$--$P_2$ exhibits the strongest direct correlation, whereas many other threshold pairs remain weakly correlated or near zero across the $R$ range. Simulations were performed for $10^6$ time steps, the first 25\% of each trajectory was discarded before analysis, and results are shown for $ENS=10$ ensemble realizations.
}
\label{fig:supp_rawcorr}
\end{figure*}

\subsection*{S7. Renewal test / aging experiment and inter-event autocorrelation}

To assess whether the events extracted from an adaptive threshold are consistent with renewal-like dynamics, we used the stripe-crossing event sequence underlying MDEA. The renewal-support analyses reported in Figs.~\ref{fig:supp_renewal_test} and~\ref{fig:supp_autotau} were performed on the information-sharing threshold of Predator 1, $I_1$, as a representative adaptive threshold. These analyses were not averaged over all six thresholds and were not computed directly from the representative CS pairs; they were used to provide mechanistic support for the event-driven interpretation of MDEA-based CS.

For each analyzed window, $I_1$ was first shifted and normalized to the interval $[0,1]$. The normalized trajectory was discretized into stripes of size $0.1$, $0.01$, or $0.001$. An event was registered whenever the signal crossed from one stripe to another. The inter-event interval sequence
\[
\{\tau_i\}
\]
was then defined as the number of samples between successive stripe-crossing events.

For the renewal aging analysis in Fig.~\ref{fig:supp_renewal_test}, the first half of the simulation was discarded, and one window of length $10^4$ samples was selected from the remaining $I_1$ trajectory. The aging time was fixed at
\[
t_a = 100.
\]
The ordinary survival function $\Psi(\tau)$ was computed from the inter-event intervals. The aged survival function was then obtained by applying the aging procedure with $t_a=100$ to the same event sequence. A shuffled-aged control was generated by randomly permuting the inter-event intervals before applying the same aging procedure. This shuffled control preserves the one-point distribution of waiting times while removing serial dependence between consecutive intervals.

The key idea of the renewal test is that, for a genuine renewal process, consecutive waiting times are statistically independent. Therefore, if the aged survival function from the original sequence and the shuffled-aged survival function are similar, the temporal ordering of the intervals carries little additional memory, which is consistent with renewal-like dynamics. Conversely, a discrepancy between the original aged and shuffled-aged curves indicates memory across successive events. The dashed black line in Fig.~\ref{fig:supp_renewal_test} is a power-law fit to the survival function over the displayed fitting range, and the reported value $\mu-1$ summarizes the estimated survival-function scaling. The aging procedure is schematized in Supplementary Fig.~\ref{fig:renewaltest}.

\begin{figure*}[tbp]
\centering
\includegraphics[width=0.8\textwidth]{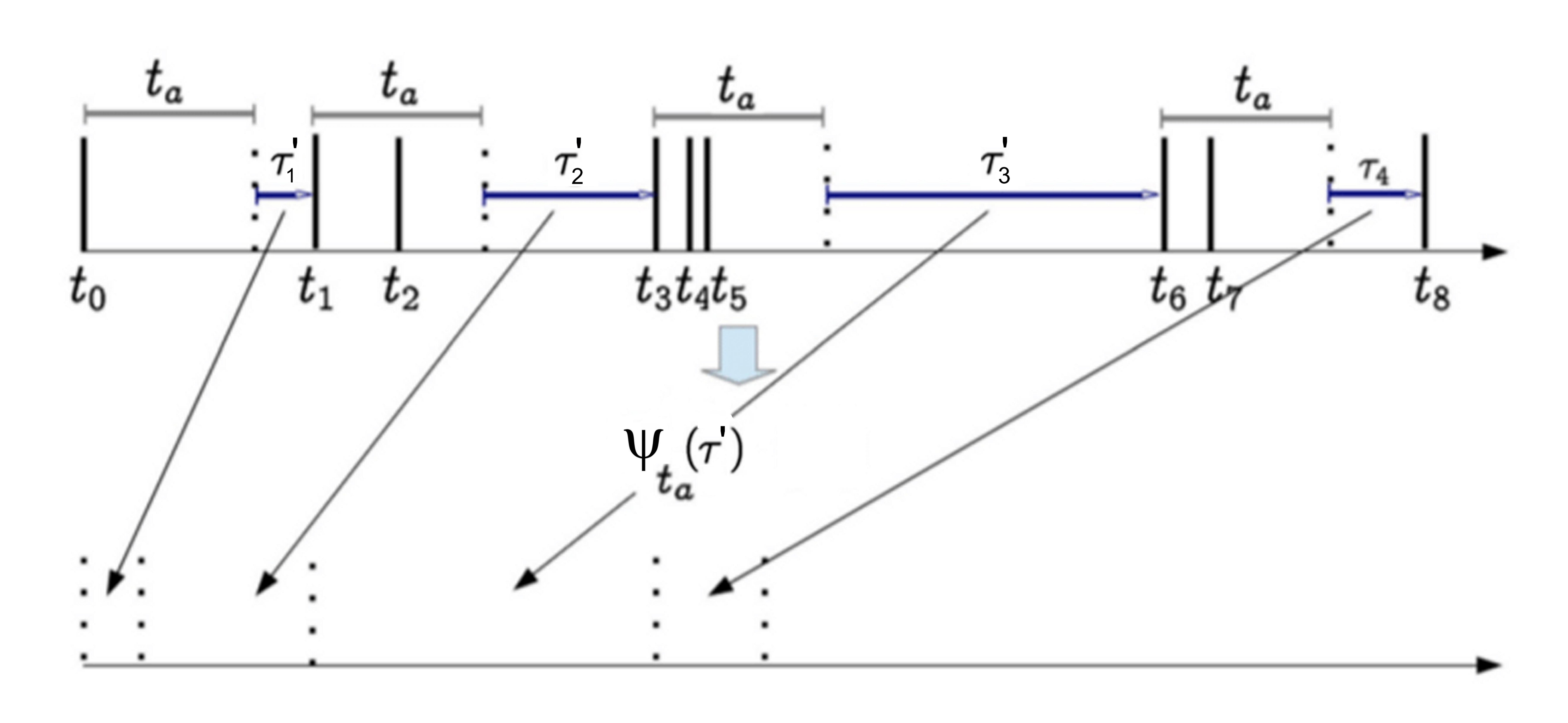}
\caption{\textbf{Supplementary Fig. S7. Schematics for the aging experiment.} Events $t_i$ are extracted using the stripe-crossing method, and the sequence is aged by an aging time $t_a$. Adapted from Ref.~14 with permission.}
\label{fig:renewaltest}
\end{figure*}

For the inter-event autocorrelation analysis in Fig.~\ref{fig:supp_autotau}, the first half of the simulation was again discarded. Ten windows of length $10^4$ samples were then selected from the remaining $I_1$ trajectory using 75\% overlap. For each window and stripe size, the inter-event interval sequence $\{\tau_i\}$ was extracted and its normalized autocorrelation was computed up to lag 100. The plotted autocorrelation curve is the mean across the 10 windows, and the shaded region represents the standard deviation across windows.

Because the autocorrelation can oscillate around zero, the memory-decay summary was obtained from the upper envelope of the absolute autocorrelation. Local peaks of $|ACF(\ell)|$ were identified as a function of lag $\ell$, and a power-law fit was performed on log--log axes. The red dashed line in Fig.~\ref{fig:supp_autotau} shows this envelope fit. The reported envelope slope quantifies the decay of inter-event memory: more negative slopes correspond to faster decay, whereas slopes closer to zero indicate more persistent correlations across event intervals.

In this way, the aging experiment and the inter-event autocorrelation provide complementary tests. The aging comparison evaluates whether temporal ordering of waiting times matters, while the autocorrelation directly quantifies memory across successive inter-event intervals. Both analyses are consistent with the interpretation that the high-cooperation regime is closer to renewal-like event organization than the low-cooperation regime.

\section*{Code availability}
All the codes used to produce the results of this work are available at \\
\url{https://github.com/Korosh137/Complexity-synchronization-as-a-diagnostic-and-control-principle}.

\section*{Acknowledgments}
Research was sponsored by the Army Research Laboratory and was accomplished under Cooperative Agreement Number W911NF-23-2-0162. The views and conclusions contained in this document are those of the authors and should not be interpreted as representing the official policies, either expressed or implied, of the Army Research Laboratory or the U.S. Government. The U.S. Government is authorized to reproduce and distribute reprints for Government purposes notwithstanding any copyright notation herein.

\section*{Author contributions}

K.M. conceived the study, developed the model and analyses, and drafted the manuscript. All authors contributed to interpretation of the results, provided critical feedback, revised the manuscript, and approved the final version.

\section*{Competing interests}
The authors declare no competing interests.

\bibliography{sample}

\end{document}